\documentclass[11pt]{article}

\usepackage[margin=1in]{geometry}

\usepackage{amsmath}
\usepackage{amssymb}
\usepackage{amsthm}
\usepackage{mathtools}

\usepackage{graphicx}
\usepackage{booktabs}
\usepackage{multirow}
\usepackage{array}

\usepackage{algorithm}
\usepackage{algpseudocode}

\usepackage{natbib}
\bibpunct[, ]{(}{)}{,}{a}{}{,}

\usepackage{microtype}
\usepackage{setspace}
\usepackage{comment}

\usepackage[hidelinks]{hyperref}
\newcommand{\Halmos}

\newtheorem{theorem}{Theorem}
\newtheorem{lemma}{Lemma}

\newtheorem{remark}{Remark}

\begin{document}

%%%%%%%%%%%%%%%%%%%%%%%%%%%%%%%%%%%%%%%%%%%%%%%%%%%%%%%%%%%%%%%%%%%%%%%%%%%%
%% Title and Authors
%%%%%%%%%%%%%%%%%%%%%%%%%%%%%%%%%%%%%%%%%%%%%%%%%%%%%%%%%%%%%%%%%%%%%%%%%%%%

\title{
Beyond Scaling: Calculable Error Bound of the Power-of-Two-Choices
Mean-Field Model in Heavy-Traffic
}

\author{
Hairi
\thanks{Department of Computer Science, University of Wisconsin-Whitewater, Whitewater, WI 53190. 
Email: \texttt{hairif@uww.edu}}
\and
Xin Liu
\thanks{Electrical Engineering and Computer Science Department, University of Michigan, Ann Arbor, MI 48109. 
Email: \texttt{xinliuee@umich.edu}}
\and
Lei Ying
\thanks{Electrical Engineering and Computer Science Department, University of Michigan, Ann Arbor, MI 48109. 
Email: \texttt{leiying@umich.edu}}
}

\date{}

\maketitle

%%%%%%%%%%%%%%%%%%%%%%%%%%%%%%%%%%%%%%%%%%%%%%%%%%%%%%%%%%%%%%%%%%%%%%%%%%%%
%% Abstract
%%%%%%%%%%%%%%%%%%%%%%%%%%%%%%%%%%%%%%%%%%%%%%%%%%%%%%%%%%%%%%%%%%%%%%%%%%%%

\begin{abstract}
This paper provides a recipe for deriving calculable approximation errors
of mean-field models in heavy-traffic with the focus on the well-known load
balancing algorithm---power-of-two-choices (Po2). The recipe combines Stein's
method for linearized mean-field models and State Space Concentration (SSC)
based on geometric tail bounds.
In particular, we divide the state space into two regions, a neighborhood near
the mean-field equilibrium and the complement of that. We first use a tail bound
to show that the steady-state probability being outside the neighborhood is
small. Then, we use a linearized mean-field model and Stein's method to
characterize the generator difference, which provides the dominant term of the
approximation error.
From the dominant term, we are able to obtain an asymptotically-tight bound,
a calculable bound, not order-wise scaling results like most results in the
literature. Finally, we compare the theoretical bound with numerical
evaluations to show the effectiveness of our results. We note that the
simulation results show that the bound is valid even for small size systems
such as a system with only hundred servers.
\end{abstract}


%%%%%%%%%%%%%%%%%%%%%%%%%%%%%%%%%%%%%%%%%%%%%%%%%%%%%%%%%%%%%%%%%%%%%%%%%%%%
%% Main text
%%%%%%%%%%%%%%%%%%%%%%%%%%%%%%%%%%%%%%%%%%%%%%%%%%%%%%%%%%%%%%%%%%%%%%%%%%%%

\section{introduction}
Large-scale and complex stochastic systems have become ubiquitous, including large-scale data centers,  the Internet of Things, and city-wide ride-hailing systems. Queueing theory has been a fundamental mathematical tool to model large-scale stochastic systems, and to analyze their steady-state performance. For example, steady-state analysis of load balancing algorithms in many-server systems is one of the most fundamental and widely-studied problems in queueing theory \citep{harchol-balter_2013}. When the stationary (steady-state) distribution is known, the mean queue length and waiting time can be easily calculated, which reveal the performance of the system and can be used to guide the design of load balancing algorithms. However, for large-scale stochastic systems in general, it is extremely challenging (if not impossible) to characterize a system's stationary distribution due to the curse of dimensionality.
For example, in a queueing system with $N$ servers, each with a buffer of size $b,$ the size of state space is at the order of $N^{b}$. Moreover, in such a system, the transition rate from a state to another may be state-dependent, so it becomes almost impossible to characterize its stationary distribution unless the system has some special properties such as when the stationary distribution has a product form \citep{SriYin_14}. 
To address these challenges, approximation methods such as mean-field models, fluid models, or diffusion models have been developed to study the stationary distributions of large-scale stochastic systems. 

This paper focuses on stochastic systems with many agents, in particular, a many-server, many-queue system such as a large-scale data center. For such systems, mean-field models have been successfully used to approximate the stationary distribution of the system in the large-system limit (when the system size becomes infinity), see e.g. the seminal papers on power-of-two-choices \citep{Mit_96,VveDobKar_96}. These earlier results, however, are asymptotic in nature by showing that the stationary distribution of the stochastic system weakly converges to the equilibrium point of the corresponding mean-field model as the system size becomes infinity. So these results do not provide the rate of convergence or the approximation error for finite-size stochastic systems. Furthermore, because the asymptotic nature of the traditional mean-field model, it only applies to the light-traffic regime where the normalized load (load per server) is strictly less than the per-server capacity in the limit. 

Both issues have been recently addressed using Stein's method. \citep{Yin_16} studied the approximation error of the mean-field models in the light-traffic regime using Stein's method, and showed that for a large-class of mean-field models, the mean-square error is $O\left(\frac{1}{N}\right),$ where $N$ is the number of agents in the system. \citep{Yin_17} further extended Stein's method to mean-field models for heavy-traffic systems and developed a framework for quantifying the approximation errors by connecting them to the local and global convergence of the mean-field models. While these results overcome the weakness of the traditional mean-field analysis, they only provide order-wise results, i.e. the scaling of the approximation errors in terms of the system size. Later, a refined mean-field analysis was developed in \citep{Gas_17,Gas_18}. In particular, \citep{Gas_18} established the coefficient of  the $\frac{1}{N}$ approximation error for light-traffic mean-field models, which provides an asymptotically exact characterization of the approximation error. However, the refined result in \citep{Gas_18} is based on the light-traffic mean-field model and the analysis uses a limiting approach. Therefore, the result and method do not apply to heavy-traffic mean-field models.

This paper obtains calculable error bound of heavy-traffic mean-field models. We consider the supermarket model \citep{Mit_96}, and assume jobs are allocated to servers according to a load balancing algorithm called power-of-two-choices \citep{Mit_96,VveDobKar_96}. While the approximation error of this system has been studied  in \citep{Yin_17}, it only characterizes the order of the error (in terms of the number of servers) and does not provide a calculable error bound. The difficulty is that the mean-field model of power-of-two-choices is a nonlinear system, so quantifying the convergence rate explicitly is difficult.  In this paper, we overcome this difficulty by focusing on a linearized heavy-traffic mean-field model, linearized around its equilibirium point, so that we can explicitly solve Stein's equation. The linearized model cannot approximate the system well when the state of the system is not near equilibrium, which is taken care of  by using a geometric tail bound to show that such deviation only occurs with a small probability. The main results of this paper are summarized below. 
\begin{itemize}
    \item For the supermarket model with power-of-two-choices, we obtain an error bound for the system in heavy-traffic. We characterize the dominating term in the approximation error, which is a function of the Jacobian matrix of the mean-field model at its equilibrium and is asymptotically accurate, which has been empirically observed to hold for hundreds of servers under heavy-traffic. We obtain the explicit forms of the bound so it is calculable given the load and the system size. 
    
    \item From the methodology perspective, the combination of state-space-concentration (SSC) and the linearized mean-field model provides a recipe for studying other mean-field models in heavy-traffic. The most difficult part of applying Stein's method for mean-field models is to establish the derivative bounds. While perturbation theory \citep{Kha_01} provides a principled approach, we can only obtain order-wise results when facing nonlinear mean-field models (see e.g. \citep{Yin_17}). Our approach is based on a basic hypothesis that if the mean-field solution would well approximate the steady-state of the stochastic system, then the steady-state should concentrate around the equilibrium point. Therefore, the focus should be on the mean-field system around its equilibrium point, which can be reduced to a linearized version. This basic hypothesis can be supported analytically using SSC based on the geometric tail bound \citep{BerGamTsi_01}. In other words, the state-space-concentration result leads to a linear system with a ``solvable'' Stein's equation, which is the key for applying Stein's method for steady-state approximation.    
\end{itemize}

\section{Related Work}
This section summarizes the related results in two categories. From the methodology perspective, this paper follows the line of research on using Stein's method for steady-state approximation of queueing systems introduced in \citep{BraDaiFen_15,BraDai_17}. This paper uses Stein's method for mean-field approximations, which has been introduced in \citep{Yin_16} and extended in \citep{Yin_17,LiuYin_18_Infocom,Gas_17,Gas_18}. Stein's method for mean-field models (or fluid models) can also be interpreted as drift analysis based on integral Lyapunov functions, which was introduced in an earlier paper \citep{Sto_15_2}. The combination of SSC and Stein's method was used in \citep{BraDai_17}, which introduces Stein's method for steady-state diffusion approximation of queueing systems. The framework is later applied to mean-field (fluid) models where Stein's equation for a simplified one-dimensional mean-field models can be solved \citep{LiuYin_19,LiuYin_20}. In this paper, the linearized system is still a multi-dimensional system. SSC  has been used in heavy-traffic analysis based on the Lyapunov-drift method, which was developed in \citep{ErySri_12} and used for analyzing computer systems and communication systems (see e.g. \citep{MagSri_16,WanMagSri_18}). 

From the perspective of the power-of-two-choices load-balancing algorithm, for the light-traffic regime,  \citep{VveDobKar_96} proved  the weak convergence of the stationary distribution of power-of-two-choices to its mean field limit in the light traffic regime, the order-wise rate of convergence was established in \citep{Yin_16}, and \citep{Gas_18} proposed a refined mean-field model with significantly smaller approximation errors.   The scaling of queue lengths of power-of-two-choices in heavy-traffic has only recently studied,  first in \citep{EscGam_16} for finite-time analysis (transient analysis) and then in \citep{Yin_17} for steady-state analysis. Our result was inspired by \citep{Gas_18}, which refines the mean-field model using the Jacobian matrix of the light-traffic mean-field equilibrium. Different from \citep{Gas_18}, based on state-space-concentration and linearized mean-field model, we established calculable error bound for heavy-traffic mean-field models where the mean-field equilibrium and the associated Jacobian matrix are both functions of the system load and system size, which prevented us from using the asymptotic approach used in \citep{Gas_18}.

\section{System Model}
In this section, we first introduce the well-known supermarket model under the power-of-two-choices load balancing algorithm. Our focus is the stationary distribution of such a system in the heavy-traffic regime (i.e. the load approaches to one as the number of servers increases). 

Then, we present the mean-field model, tailored for the $N$-server system \citep{Yin_17} and the exact load of the $N$-server system. The solution of the mean-field model is an  approximation of the stationary distribution of the stochastic system. We will then present the approach to characterize the approximation error based on Stein's equation.

Consider a many-server system with $N$ homogeneous servers, where job arrivals follow a Poisson process with rate $\lambda N$ and service times are i.i.d. exponential random variables with rate one. Each server can hold at most $b$ jobs, including the one in service.  We consider $\lambda=1-\frac{\gamma}{N^{\alpha}}$ for some $0<\gamma\le 1$ and $\alpha \geq 0$. When $\alpha=0$, $\lambda$ is  a constant independent of $N$ which we call the light-traffic regime. When $\alpha>0$, the arrival rate depends on $N$ and approaches to one as $N\rightarrow \infty,$ which we call the heavy-traffic regime. We assume the system is under a load balancing algorithm called power-of-two-choices \citep{Mit_96,VveDobKar_96}.

\noindent {\bf Power-of-Two-Choices (Po2):}
Po2 samples two servers uniformly at random among $N$ servers and dispatches the incoming job to the server with the shorter queue size. Ties are broken uniformly at random.

\begin{figure}[htbp]
\centerline{\includegraphics[width=3.in]{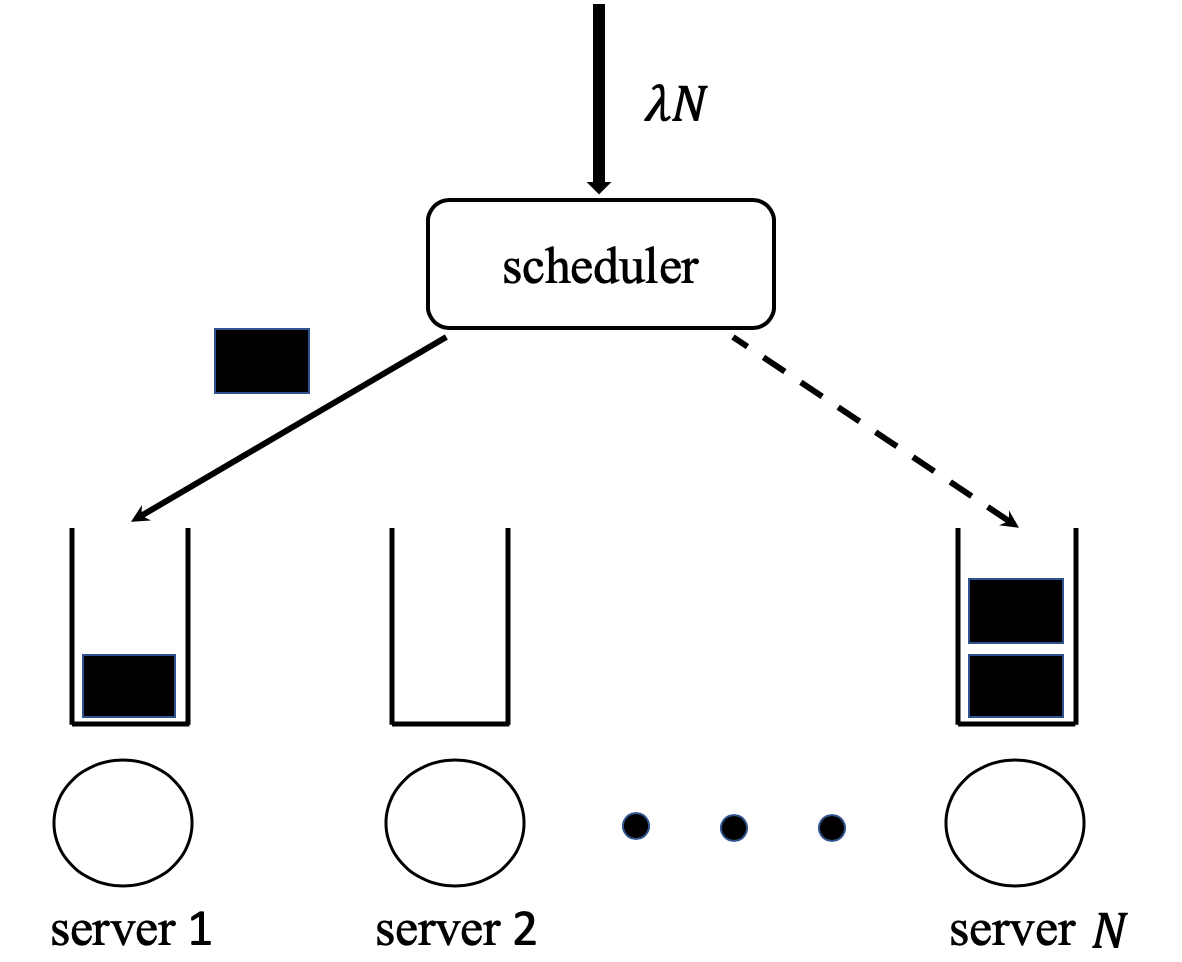}}
  \caption{Power-of-Two-Choices}
  \label{Po2}
\end{figure}

Let $S_i(t)$ denote the fraction of servers with queue size at least $i$ at time $t.$ The term $S_0(t)=1, \forall t$ by definition. Under the finite buffer assumption with buffer size $b$, $S_i(t)=0, \forall i\ge b+1, \forall t.$ Throughout the paper, we assume that the buffer size b can be up to the order of $\log N$, i.e. $b=O(\log N)$.
Define $\mathcal{S}$ to be
\begin{align}
\mathcal{S}=\{s ~|~ 1\ge s_1\ge \cdots \ge s_b\ge 0\}, \nonumber
\end{align}
and $S(t)=[S_1(t),S_2(t),\cdots,S_b(t)]$. It is easy to verify that the state $S(t)$ is a continuous time Markov chain (CTMC). Define $e_k$ to be a $b$-dimensional vector such that the $k$th entry is one and all other entries are zero. We characterize the transition rates $R_{s,s'}$ from $s$ and $s'$ as follows:

\[R_{s,s'}= \\
\begin{cases} 
      N(s_k-s_{k+1}),  \text{if } s'=s-\frac{e_k}{N} \text{ and } 1\le k\le b-1\\
      Ns_b, \text{if } s'=s-\frac{e_b}{N} \\
      \lambda N(s^{2}_{k-1}-s^{2}_k), \text{if } s'=s+\frac{e_k}{N} \\
      \sum_{k=1}^{b}-\lambda N(s^{2}_{k-1}-s^{2}_k)-N(s_k-s_{k+1}),  \text{if } s'=s \\
      0, \text{otherwise} 
   \end{cases}\]
The first and second terms correspond to the event that a job departs from a server with queue size $k$ so $s_k$ decreases by $\frac{1}{N},$ and the third term corresponds to the event that a job arrives and joins a server with queue size $k-1.$ We define a normalized transition rate to be 
\begin{align}
q_{s,s'}=\frac{R_{s,s'}}{N}. \nonumber
\end{align}

We focus on the steady-state analysis of the system, i.e. the distribution of $S(\infty)$. At the steady-state, $S(\infty)$ is a $b$-dimensional random vector. For simplicity, we let $S$ denote $S(\infty)$. In this paper, we use uppercase letters for random variables and lowercase letters for deterministic values.  

The mean-field model \citep{Mit_96,VveDobKar_96,Yin_17} for this system is 
\begin{align}
\dot{s}=f(s)=\sum_{s':s'\neq s}R_{s,s'}(s'-s)=N\sum_{s':s'\neq s}q_{s,s'}(s'-s). \nonumber
\end{align}
According to the definition of $R_{s,s'}$ and $q_{s,s'},$ we have
\[ \dot{s}_k=f_k(s)=\begin{cases} 
      \lambda(s^{2}_{k-1}-s^{2}_k)-(s_k-s_{k+1}), & 1 \leq k \leq b-1 \\
      \lambda(s^{2}_{b-1}-s^{2}_b)-s_b, & k=b. 
   \end{cases}
\]
The equilibrium point of this mean-field model, denoted by $s^{*},$  satisfies the following conditions:
\begin{subequations}
\begin{align}
s^{*}_0&=1 \\
\lambda\left((s^{*}_{k-1})^{2}-(s^{*}_k)^{2}\right)-(s^{*}_k-s^{*}_{k+1})&=0, \quad 1\le k\le b-1 \label{eq: 1_b}\\
\lambda\left((s^{*}_{b-1})^{2}-(s^{*}_b)^{2}\right)-s^{*}_b&=0.  
\end{align}
\label{eq: equ}
\end{subequations}
The existence and uniqueness of the equilibrium point has been proved in \citep{Mit_96}. Define 
\begin{align}
g(s)=-\int_{0}^{\infty}d(s(t),s^{*})dt, \quad s(0)=s. \nonumber
\end{align}
where $d(s(t),s^{*})$ is a distance function. Then, by the definition of $g(s),$ we have
\begin{align}
\nabla g(s)\cdot f(s)=d(s,s^{*}). \label{PossionEq}
\end{align}
Equation \eqref{PossionEq} is called the Poisson equation or Stein's equation.
For any bounded $g,$ we have the following steady state equation (Basic Adjoint Relationship (BAR) \citep{GlyZee_08})
\begin{align}
\mathbb E[Gg(S)]=0, \label{BAR}
\end{align}
where the expectation is taken with respect to the steady state distribution of $S$ and $G$ is the generator of the CTMC. Combining \eqref{PossionEq} and \eqref{BAR}, we have
\begin{align}
\mathbb E\left[d(S,s^{*})\right]
=&\mathbb E\left[\nabla g(S)\cdot f(S) - Gg(S)\right]\nonumber\\
=&-\mathbb E\left[\sum_{s'}R_{S,s'}\Gamma(S,s')\right], \label{dist-gen} 
\end{align}
where $\Gamma(s,s')=g(s')-g(s)-\nabla g(s)\cdot(s'-s)$. 
From \eqref{dist-gen}, Stein's method provides us a way to study the approximation error defined by $\mathbb E[d(S, s^*)]$ by bounding the generator difference between the original system and the mean-field model.

\section{Main Results and Methodology}
This section summarizes our main results, which include an asymptotically tight approximation error bound. We remark again the bound in Theorem \ref{thm:heavy traffic convergence} can be calculated numerically and is not order-wise result as in most earlier papers. 

\begin{theorem}[Asymptotically Tight Bound]\label{thm:convergence}
For $0<\alpha<\frac{1}{18}$, we have that 
\begin{align}
\mathbb E[||S-s^{*}||^{2}]=\frac{2}{N}\text{vec}(\tilde{F})^{\top}(J^{\top}(s^{*})\otimes I+ I\otimes J^{\top}(s^{*}))^{-1} \text{vec}(I)+o\left(\frac{1}{N^{1+\alpha}}\right)
\end{align}
where $\text{vec}(\cdot)$ is vectorization of a matrix, $\otimes$ Kronecker product, $I$ identity matrix,
\begin{equation}
J(s^*)=\left[ \begin{matrix}
-2\lambda s^*_1-1 & 1 &  & 0\\
2\lambda s^*_1 & \ddots  & \ddots & \\
 & \ddots & \ddots & 1 \\
0 & &2\lambda s^*_{b-1} & -2\lambda s^*_b-1 \end{matrix} \right] \nonumber
\end{equation} is the Jacobian matrix of the mean-field model $f(s)$ at equilibrium point $s^{*},$ and $\tilde{F}:= \text{diag}(\tilde{f}_1(s^{*}),\cdots,\tilde{f}_b(s^{*}))$ with
$$\tilde{f}_i(s^{*})=\frac{1}{2}[\lambda\left((s^{*}_{i-1})^{2}-(s^{*}_i)^{2}\right)+(s^{*}_i-s^{*}_{i+1})]$$ for $i=1, 2, \cdots, b$.
\label{thm:heavy traffic convergence}
\Halmos
\end{theorem}
The theorem states that the mean square error $\mathbb E[||S-s^{*}||^{2}]$ has an asymptotic dominating term $$\frac{2}{N}\text{vec}(\tilde{F})^{\top}(J^{\top}(s^{*})\otimes I+ I\otimes J^{\top}(s^{*}))^{-1} \text{vec}(I).$$ Therefore, we have
\begin{align}
\lim_{N\rightarrow \infty } N\mathbb E[||S-s^{*}||^{2}]=\frac{2}{N}\text{vec}(\tilde{F})^{\top}(J^{\top}(s^{*})\otimes I+ I\otimes J^{\top}(s^{*}))^{-1} \text{vec}(I). \label{eq: dom_term}
\end{align} 

The invertibility of $J^{\top}(s^{*})\otimes I+I\otimes J^{\top}(s^{*})$ is guaranteed in Lemma \ref{lem: Kro_inv}.
In general, to obtain the dominant term \eqref{eq: dom_term}, one has to first compute a positive definite matrix $P$ in its defined form or by solving a related Lyapunov equation \citep{Gas_18}, then numerically calculate the dominant term expression even in the light-traffic regime. In the above theorem, we provide that the dominant term has a closed form solution, thus a calculable error bound.

\begin{remark}[General Upper Bound]
For $0<\alpha<\frac{1}{18}$ and any arbitrarily small $\eta>0$, there's a sufficiently large $N$, such that 
\begin{align}
\mathbb E[||S-s^{*}||^{2}]\le\frac{2+\eta}{N}\text{vec}(\tilde{F})^{\top}(J^{\top}(s^{*})\otimes I+ I\otimes J^{\top}(s^{*}))^{-1} \text{vec}(I). \nonumber
\end{align}
\label{cor: gup}
\end{remark}
This upper bound result easily follows from the asymptotic tight bound result.
%This result tells us that we can have a calculable upper bound for heavy-traffic which holds for finite $N$.

 \begin{remark}[Order-wise Convergence]
 For arbitrarily small $\xi>0$, there is a sufficiently large $N$ such that for $0<\alpha<\frac{1}{18}$ the following holds
\begin{equation}
    \mathbb E[||S-s^{*}||^{2}] \le \frac{1}{N^{1-2\alpha-4\xi}}. \nonumber
\end{equation}
 
% \[ \mathbb E[||S-s^{*}||^{2}]\le\begin{cases} 
%       \frac{1}{N^{1-2\alpha-4\xi}}, & 0<\alpha<\frac{1}{12} \\
%       \frac{1}{N^{1-4\alpha-7\xi}}, & \frac{1}{12}\le\alpha<\frac{1}{4}. 
%   \end{cases}
% \] 
 \label{thm: order-wise}
 \end{remark}
This result characterizes an order-wise upper bound on mean square error for $\alpha\in (0,\frac{1}{18})$. This is a straightforward result from upper bound analysis of the dominant term in Lemma \ref{lem: 2nd_upp_bou}.

\begin{figure}[htbp]
  \centering
  \includegraphics[width=3.in]{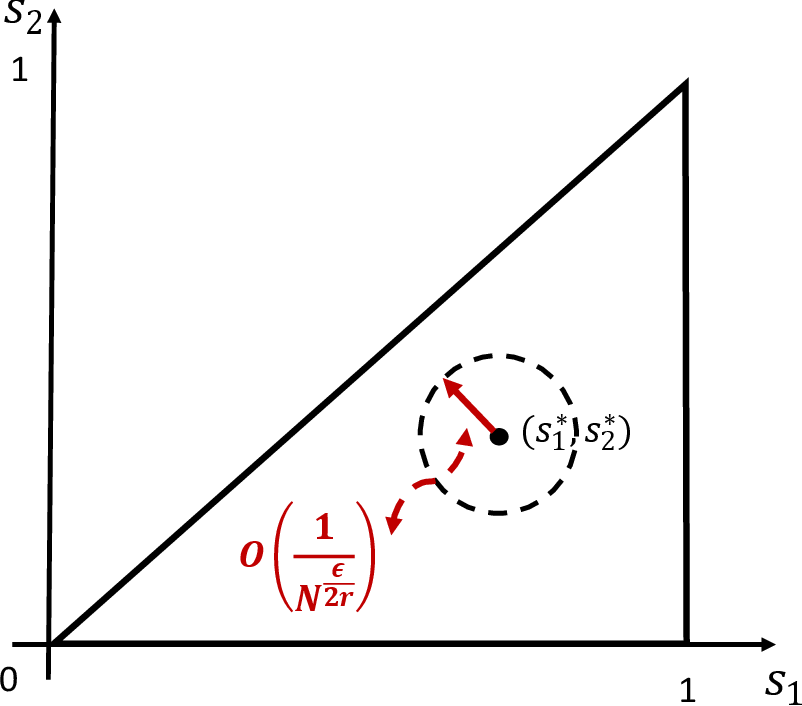}
  \caption{$b=2$ Illustration of inside region and outside region}
  \label{illustration}
\end{figure}

Our analysis combines Stein's method with a linear dynamical system and SSC. We divide the state space into two regions based on the mean-field solution; specifically, one region including the states that are ``close'' to the equilibrium point $s^*,$ and the other region that includes all other states. 
For example, consider the case of $b=2$, so the state space is two-dimensional as shown in Figure \ref{illustration}, where the state space is $$\left\{\left.(s_1,s_2)=\left(\frac{p}{N},\frac{q}{N}\right)\right|p\geq q\in\{0,1,\cdots,N\}\right\}\subset [0,1]^{2}.$$ As in Figure \ref{illustration}, we divide the state space into two regions separated by the dashed circle. The size of the circle is small and depends on $N,$ in particular, the radius is $O\left(\frac{1}{N^{\frac{\epsilon}{2r}}}\right),$ where both $\epsilon$ and $r$ are positive values (the choices of these two values become clear in the analysis). 

For the two different regions, we apply different techniques: 
\begin{enumerate}
    \item We first establish higher moment bounds that upper bound the probability that the steady state is outside the dashed circle.  The proof is based on the geometric tail bound in \citep{Haj_82,BerGamTsi_01} and by showing that there is a ``significant'' negative drift that moves the system closer to the equilibrium point when the system is outside of the dashed circle. 
    \item For the states close to the equilibrium point, i.e, inside the dashed circle, from the control theory, we know that the mean-field nonlinear system behavior can be well approximated by the linearized dynamical system. By carefully choosing the parameters, we can look into the generator difference and calculate the dominant term of the approximation error by using the linearized mean-field model. The linearity enables us to solve Stein's equation, which is a key obstacle in applying Stein's method. 
\end{enumerate}

\section{Simulations}
Given $\alpha=0.05$, we performed simulations for two different choices of $\gamma$ and different system sizes. The purpose of these simulations is to compare the approximation errors calculated from the simulations with the asymptotically tight bound. The results are based on the average of 10 runs, where each run simulates $10^{9}$ time steps. We averaged over the last $9\times 10^{8}$ time slots of each run  to compute the steady state values. 

For each run, we calculated the empirical mean square error multiplied by the system size $N.$ 
%since our theoretical analysis suggests that $NE[\|S-s^*\|^2]$ does not scale with $N.$ 
Recall that the asymptotically tight bound is $$2\text{vec}(\tilde{F})^{\top}(J^{\top}(s^{*})\otimes I+ I\otimes J^{\top}(s^{*}))^{-1} \text{vec}(I).$$ 

\begin{table}[htbp]
\caption{$\gamma=0.1, \alpha=0.05$}
\begin{center}
\begin{tabular}{|c|c|c|c|c|}
\hline
$N$& 10 & 100 & 1,000 & 10,000 \\
\hline
$\lambda$& 0.9109 & 0.9206 & 0.9292 & 0.9369 \\
\hline
Simulation & 4.2975 & 3.6884 & 3.9553 & 4.4068 \\
\hline
Asymptotic Tight Bound & 3.1872 & 3.5499 & 3.9542 & 4.4049 \\
\hline
%Old Bound & 3.2773 & 3.6411 & 4.0455 & 4.4955 \\
%\hline
\end{tabular}
\label{gamma0_1}
\end{center}
\end{table}

\begin{table}[htbp]
\caption{$\gamma=0.01, \alpha=0.05$}
\begin{center}
\begin{tabular}{|c|c|c|c|c|}
\hline
$N$& 10 & 100 & 1,000 & 10,000 \\
\hline
$\lambda$& 0.9911 & 0.9921 & 0.9929 & 0.9937 \\
\hline
Simulation & 77.9532 & 46.4641 & 38.7702 & 40.2093 \\
\hline
Asymptotic Tight Bound & 28.8070 & 32.2472 & 36.1022 & 40.4250 \\
\hline
%Old Bound & 28.2972 & 31.6293 & 35.3629 & 39.5457 \\
%\hline
\end{tabular}
\label{gamma0_01}
\end{center}
\end{table}

Tables \ref{gamma0_1} and \ref{gamma0_01} summarize the results with $\gamma=0.1, \alpha=0.05$ and $\gamma=0.01, \alpha=0.05$ respectively. We varied the size of the system in both cases. Note that the arrival rate is a function of the system size and approaches one as $N$ increases. The simulation results are in the same order with the asymptotic tight bounds even for smaller system sizes.
%the dominant terms and are bounded by the upper bounds. 
%Interestingly, the empirical values are in the middle of the asymptotic bound and the upper bound in all the cases. 

Our numerical results show that the asymptotic bound matches the empirical error very well, and approaches the empirical error as $N$ increases. In particular, for $\gamma=0.1$ and $\alpha=0.05,$ the results are close even when $N=100;$ and for $\gamma=0.01$ and $\alpha=0.05,$ the results are close when $N=1,000.$ 

%As we can see, the upper bound is valid even for small size systems, e.g. $N=10$, which shows the effectiveness of our results. 
From a practical point of view, the asymptotic tight bounds are calculable, so they provide good estimates of the mean-square error.

\section{Proofs}

\begin{figure}[htbp]
  \centering
  \includegraphics[width=.8\linewidth]{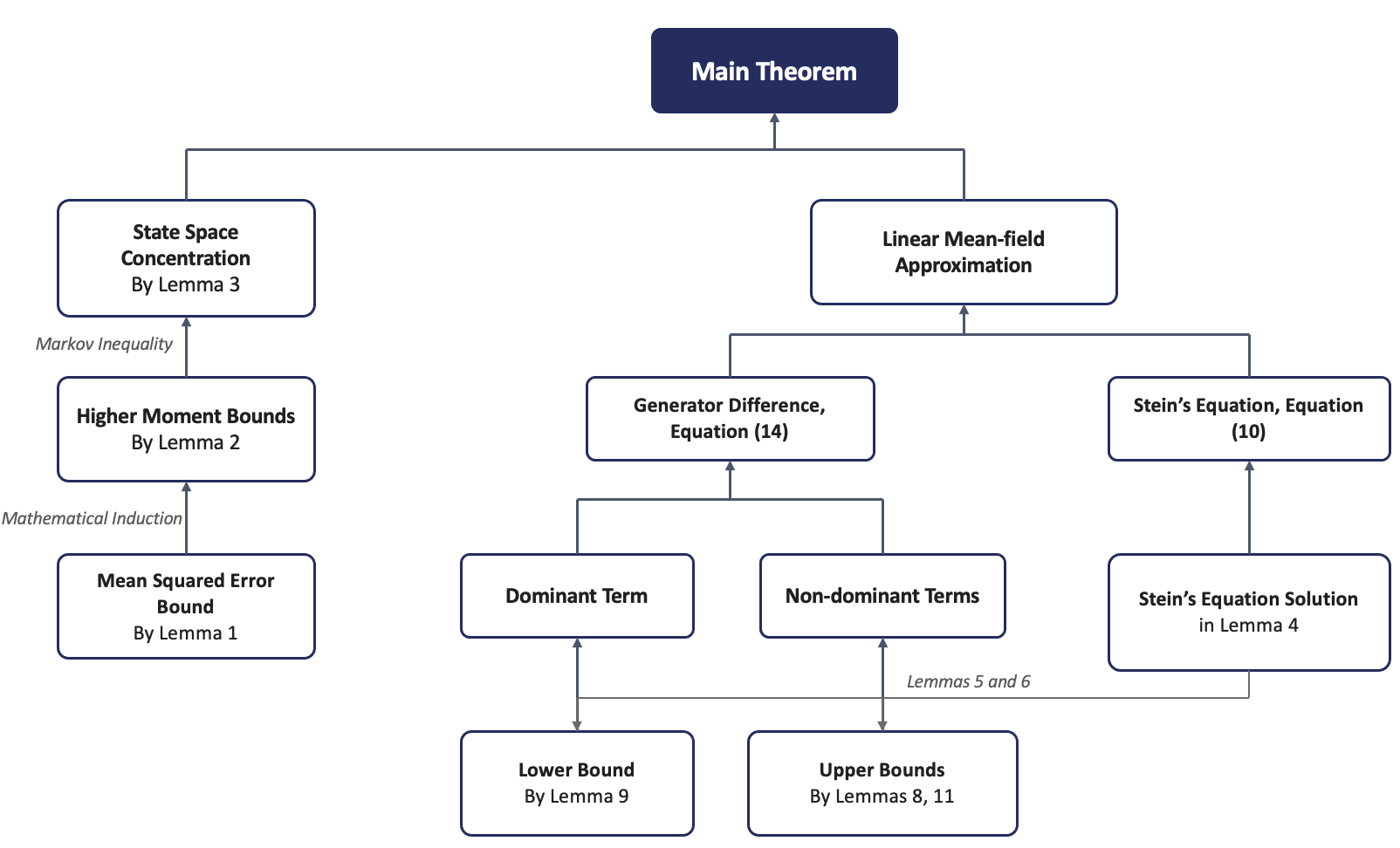}
  \caption{flowchart of proofs}
  \label{fig: flowchart}
\end{figure}

As we mentioned earlier, the results are established by looking at the system in two different regions, near the equilibrium point and outside.
The flow chart of the proofs is given in Figure \ref{fig: flowchart}.

The main result consists of two components: State Space Concentration (SSC) and the Linear Mean-field Approximation.

For the SSC component, we first establish the usual first-order mean-squared error bound, following the approach in \cite{Yin_17}, as stated in Lemma \ref{lem: pre_con}. We then use mathematical induction to prove that higher-order moment bounds hold for any $r\in\mathbb{N}$, as shown in Lemma \ref{lem: hig_mom}. Finally, by applying Markov's inequality, we show that the probability of the state deviating from a ball centered at the mean-field solution with radius $\frac{1}{N^{\frac{\epsilon}{2r}}}$ is small. This establishes the state space concentration property within the specified ball.

Once the SSC result is established, we characterize the expected mean-squared error of the CTMC of Po2 to the mean-field solution $s^{*}$ by applying the linear mean-field model to approximate the stochastic system. The generator difference, given in Eq.\eqref{eq: gen_dif}, consists of three terms. We then identify the dominant and non-dominant terms by: 1) solving Stein's equation \eqref{eq: Poi_eq} using Lemma \ref{lem:Poi_sol}; and 2) deriving upper and lower bounds for the resulting solution. Specifically, we show that the lower bound of the dominant term (Lemma \ref{lem: 2nd_low_bou}) exceeds the upper bounds of the non-dominant terms (Lemmas \ref{lem: 1st_upp_bou} and \ref{lem: 3rd_upp_bou}).

In the subsequent analysis, we use $\|\cdot\|$ and $|\cdot|$ to denote $\ell_2$ and $\ell_1$ norms respectively. 
\subsection{State Space Concentration}
First, we present some preliminary heavy-traffic convergence results for finite buffer size  $b=O(\log N)$. 
\begin{lemma}\label{lem: pre_con}
For any $0<\alpha<0.25$ and arbitrarily small $\xi>0$, there is a sufficiently large $N$ that the following holds
\begin{align}
\mathbb E[||S-s^{*}||^{2}]\le\frac{1}{N^{1-4\alpha-7\xi}}. \nonumber 
\end{align}
\end{lemma}

\begin{lemma}[Higher Moment Bounds]
For $r\in\mathbb{N}$, any $0<\alpha<0.25$ and arbitrarily small $\xi>0$, there is a sufficiently large $N$ that the following holds
\begin{align}
\mathbb E\left[||S-s^{*}||^{2r}\right]\le \frac{1}{N^{r(1-4\alpha-7\xi)}}. 
\label{eq: hig_mom_bou}
\end{align}
\label{lem: hig_mom}
\end{lemma}
The proofs for both lemmas are in the appendix.

\begin{lemma}[State Space Concentration] Letting $\epsilon>0$ and $r\in\mathbb{N}$, for arbitrarily small $\xi>0$, there is a sufficiently large $N$ that the following holds

\begin{align}
\mathbb{P}\left(||S-s^{*}||^{2r}\ge\frac{1}{N^{\epsilon}}\right)\le \frac{1}{N^{r(1-4\alpha-7\xi)-\epsilon}}. \nonumber
\end{align}
\label{lem: ssc}
\end{lemma}
\proof{Proof.}
Simply applying Markov inequality to the result of Lemma \ref{lem: hig_mom}, we have 
\begin{align}
\mathbb{P}(||S-s^{*}||^{2r}\ge\frac{1}{N^{\epsilon}})&\le \frac{\mathbb{E}[||S-s^{*}||^{2r}]}{\frac{1}{N^{\epsilon}}} \nonumber\\
&\le\frac{N^{\epsilon}}{N^{r(1-4\alpha-7\xi)}} \nonumber \\
&=\frac{1}{N^{r(1-4\alpha-7\xi)-\epsilon}}.  \nonumber 
\end{align} \Halmos
\endproof

\subsection{Linear Mean-field Model}
Define a set of states to be  $\mathcal B = \{s ~ | ~ ||s-s^{*}||^{2r}\le\frac{1}{N^{\epsilon}}\},$ which are the states close to the equilibrium point. 
Let $d(s,s^{*})=||s-s^{*}||^{2}$ be the distance function. We consider a simple linear system 
\begin{align}\label{eq: linear mfm}
\dot{s}=l(s)=J(s^{*})(s-s^{*}),
\end{align}
where $J(s^{*})$ is the Jacobian matrix of $f(s)$ at the equilibrium point $s^{*}.$ In heavy-traffic, the entries of $J(s^{*})$ is generally also a function of $N$ as is $s^{*}$ itself.

The Jacobian matrix at $s$ is 
\begin{equation}
J(s)=\left[ \begin{matrix}
-2\lambda s_1-1 & 1 &  & 0\\
2\lambda s_1 & \ddots  & \ddots & \\
 & \ddots & \ddots & 1 \\
0 & &2\lambda s_{b-1} & -2\lambda s_b-1 \end{matrix} \right] \nonumber
\end{equation}

We introduce a lemma to solve Stein's equation (the Poisson equation) for the linear mean-field system.
Consider a function $g: \mathcal{S}\to\mathcal{R}$ such that it satisfies the following equation
\begin{align}
    Lg(s)\dot{=}\frac{dg(s)}{dt}=\nabla g(s)\cdot l(s)=||s-s^{*}||^{2}.
\end{align}
According to the definition of the linear mean-field model in \eqref{eq: linear mfm}, we have
\begin{align}
    \nabla g(s)\cdot J(s^{*})(s-s^{*})=||s-s^{*}||^{2}.
    \label{eq: Poi_eq}
\end{align}
\begin{lemma}[Solution of Stein's Equation]
The solution of the Poisson equation \eqref{eq: Poi_eq} satisfies
$$g(s)=(s-s^{*})^{\top}P(s-s^{*})$$
where $P$ is defined as $P= \int_{0}^{\infty}  e^{J(s^{*})^{\top}t}e^{J(s^{*})t} dt$
and furthermore
\begin{align}
\nabla g(s) &= -2P (s-s^{*}) \nonumber \\
\nabla^{2} g(s) &= -2P \nonumber \\
\nabla^{3} g(s) &= 0. \nonumber 
\end{align}
\label{lem:Poi_sol}
\end{lemma}
\proof{Proof.}
By \eqref{eq: linear mfm}, we have the solution for linear ODE as follows.
\begin{align}
s(t) - s^{*} = e^{J(s^{*})t}(s-s^{*}) , \nonumber
\end{align}
where $s$ denotes an initial state.
According to Stein's equation \eqref{eq: Poi_eq}, we have
\begin{align}
g(s) &=  -\int_{0}^{\infty} \|s(t) - s^{*}\|^{2} dt \nonumber \\
&= -\int_{0}^{\infty} \|e^{J(s^{*})t} (s- s^{*})\|^{2} dt \nonumber \\
&= -\int_{0}^{\infty} (s - s^{*})^{\top} e^{J(s^{*})^{\top}t}e^{J(s^{*})t} (s-s^{*})dt \nonumber \\
&= -(s - s^{*})^{\top} \int_{0}^{\infty}  e^{J(s^{*})^{\top}t}e^{J(s^{*})t} dt \cdot (s-s^{*}) \nonumber \\
&= -(s - s^{*})^{\top} P \cdot (s-s^{*}) \nonumber
\end{align}
where we denote $P:= \int_{0}^{\infty}  e^{J(s^{*})^{\top}t}e^{J(s^{*})t} dt$.

Then, the gradient and Hessian are
\begin{align}
\nabla g(s) &= -2P (s-s^{*}) \nonumber \\
\nabla^{2} g(s) &= -2P \nonumber \\
\nabla^{3} g(s) &= 0 \nonumber.
\end{align}
We also note that $P$ is the unique solution to the following Lyapunov equation
\begin{align}
PJ(s^{*}) + J(s^{*})^{\top}P = -I. \nonumber
\end{align}
\Halmos
\endproof

We next provide two lemmas establishing a lower bound on the diagonal entries of $P$ and an upper bound on $\|P\|$, which will be used for proofs that follows.
\begin{lemma}[Lower Bound on Diagonal Entries $P_{ii}$]
For any $i\in \{1,\cdots,b\}$, we have
\begin{align}
P_{ii}\ge \frac{1}{2\|J(s^{*})\|}\ge \frac{1}{18}. \nonumber
\end{align}
\label{lem: P_low_bound}
\end{lemma}
\proof{Proof.}
Let $e_i$ be the canonical unit basis vector. Then, the diagonal entry $P_{ii}$ is as follows
\begin{align}
P_{ii} &= e^{\top}_i P e_i \nonumber \\
&= e^{\top}_i \int_{0}^{\infty}  e^{J(s^{*})^{\top}t}e^{J(s^{*})t} dt e_i \nonumber \\
&= \int_{0}^{\infty}  e^{\top}_i e^{J(s^{*})^{\top}t}e^{J(s^{*})t}e_i dt  \nonumber \\
&= \int_{0}^{\infty}  \|e^{J(s^{*})t}e_i\|^2 dt.  \nonumber
\end{align}
For the integrand, we consider the derivative
\begin{align}
\frac{d \|e^{J(s^{*})t}e_i\|^2}{dt} &= e^{\top}_i J(s^{*})^{\top} e^{J(s^{*})^{\top}t}e^{J(s^{*})t}e_i+ e^{\top}_i e^{J(s^{*})^{\top}t}J(s^{*}) e^{J(s^{*})t}e_i \nonumber \\
&= e^{\top}_i \left [J(s^{*})^{\top} e^{J(s^{*})^{\top}t} + e^{J(s^{*})^{\top}t}J(s^{*}) \right]e^{J(s^{*})t}e_i \nonumber \\
&=_{(i)} e^{\top}_i \left [ e^{J(s^{*})^{\top}t}J(s^{*})^{\top} + e^{J(s^{*})^{\top}t}J(s^{*}) \right]e^{J(s^{*})t}e_i \nonumber \\
&= e^{\top}_i e^{J(s^{*})^{\top}t} \left [J(s^{*})^{\top} + J(s^{*}) \right] e^{J(s^{*})t}e_i. \nonumber
\end{align}
The equality $(i)$ is by commutability of exponential matrix and itself. By Cauchy-Schwarts inequalities, we have
\begin{align}
\frac{d \|e^{J(s^{*})t}e_i\|^2}{dt} &\ge - \|e^{\top}_i e^{J(s^{*})^{\top}t}\|\cdot \|J(s^{*})^{\top} + J(s^{*})\|\cdot \| e^{J(s^{*})t}e_i\| \nonumber \\
&\ge - \| e^{J(s^{*})^{t}}e_i\|^{2} (\|J(s^{*})^{\top}\| + \|J(s^{*})\|) \nonumber \\
&= -2  \| e^{J(s^{*})^{t}}e_i\|^{2} \|J(s^{*})\|. \nonumber
\end{align}
By Gr\"{o}nwall inequality, we have 
\begin{align}
\|e^{J(s^{*})t}e_i\|^2\ge e^{-2\int_0^{t} \|J(s^{*})\| d\tau} = e^{-2 \|J(s^{*})\|t}. \nonumber
\end{align}
As a result, we have
\begin{align}
P_{ii}\ge \int_0^{\infty}  e^{-2 \|J(s^{*})\|t} dt \ge \frac{1}{2\|J(s^{*})\|}. \nonumber
\end{align}
For $\|J(s^{*})\|$, we have
\begin{align}
\|J(s^{*})\|\le \sqrt{\|J(s^{*})\|_{1} \|J(s^{*})\|_{\infty}} \le 9. \nonumber
\end{align}
We used the fact that $J(s^*)$ is a tridiagonal matrix and $J(s^*)_{ij}\le 3$ for all $i,j$, which implies that $\|J(s^{*})\|_{1}\le 9, \|J(s^{*})\|_{\infty}\le 9$ respectively.
\Halmos
\endproof

\begin{lemma}[Upper Bound on $\|P\|$]
For arbitrarily small $\xi>0$, there is a sufficiently large $N$ that the following bound on the norm of $P$ holds
$$\|P\| \le N^{2\alpha+3\xi}.$$
\label{lem: p_bound}
\end{lemma}
\proof{Proof.}
To bound $\|P\|$, we have
\begin{align}
\|P\| &= \left\|\int_{0}^{\infty}  e^{J(s^{*})^{\top}t}e^{J(s^{*})t} dt\right\| \nonumber \\
&\le \int_{0}^{\infty}  \left\|e^{J(s^{*})^{\top}t}e^{J(s^{*})t}\right\| dt \nonumber \\
&= \int_{0}^{\infty} \left\|e^{J(s^{*})t}\right\|^{2}  dt. \nonumber 
\end{align}

Observe that Lemma C.5 in \cite{Yin_17} is for original nonlinear system, we use the fact that linearized system preserves the exponential convergence under the same Lyapunov function. Equivalently, one can use the exactly same argument for Linear system by replacing nonlinear system. In other words, for the linear counterpart, we also have
$$|x(t)|\le 3|x(0)|e^{-\delta_0 t},$$
where $x(t) = s(t) - s^{*}$ and $\delta_0\ge \frac{1}{N^{2\alpha+2\xi}}$ by Lemma C.4 in \cite{Yin_17}.

Also observe that for the linear system, the solution is $x(t)= e^{J(s^*)t}x(0)$. In particular, when we consider
\begin{align}
|e^{J(s^{*})t}x(0)|\le 3|x(0)|e^{-\delta_0t} \nonumber
\end{align}
means that
\begin{align}
|e^{J(s^{*})t}|:=\sup_{x(0)\neq 0}\frac{|e^{J(s^{*})t}x(0)|}{|x(0)|} \le 3e^{-\delta_0t}. \nonumber 
\end{align}
From norm equivalence, we know that 
\begin{align}
\|e^{J(s^{*})t}\|\le \sqrt{b}|e^{J(s^{*})t}| \le 3\sqrt{b}e^{-\delta_0t}. \nonumber
\end{align}
Then, plug it in 
\begin{align}
\|P\| &\le \int_{0}^{\infty}  \|e^{J(s^{*})t}\|^{2} dt \nonumber \\
&\le \int_{0}^{\infty} 9be^{-2\delta_0t} dt \nonumber \\
&= 9b\frac{1}{-2\delta_0} e^{-2\delta_0t}|_{0}^{\infty} \nonumber \\
&= 9b\frac{1}{2\delta_0} \nonumber \\
&\le 54b \frac{N^{2\alpha+2\xi}}{\gamma} \nonumber \\
&\le N^{2\alpha+3\xi}. \nonumber
\end{align}
The second from the last inequality is by the fact $\delta_0\ge \frac{1}{N^{2\alpha+2\xi}}$ and the last equality holds for sufficiently large $N$. \Halmos
\endproof

\subsection{Proof of the theorem}
We start from the mean square error by studying the generator difference when state $S$ is close to $s^*.$ In particular, we focus on 
\begin{align}
\mathbb E\left[Lg(S)-Gg(S)\left| S\in\mathcal B\right.\right].
\end{align}

\begin{lemma}
The generator applying to function $g(s)$ satisfies
\begin{align}
    Gg(s)=\nabla g(s)\cdot f(s)+\frac{1}{N}\sum_{i=1}^{b}\nabla^{2} g(s)_{ii} \tilde{f}_i(s)
\end{align}
where $\nabla^{2} g(s)_{ii}$ is the $i$-th diagonal element of the Hessian matrix $\nabla^{2} g(s)$ and 
$$\tilde{f}_i(s)=\frac{1}{2}[\lambda(s^{2}_{i-1}-s^{2}_i)+(s_i-s_{i+1})].$$ 
\end{lemma}
\proof{Proof.}
According to the definition of generator $G,$ we have 
\begin{align}
Gg(s)
=&\sum_{i=1}^{b}\lambda N(s^{2}_{i-1}-s^{2}_i)[g(s+e_i)-g(s)] +N(s_i-s_{i+1})[g(s-e_i)-g(s)]. \nonumber
\end{align}
By the Taylor expansion at the state $s,$ we have
\begin{align}
Gg(s)=&\sum_{i=1}^{b}\lambda N(s^{2}_{i-1}-s^{2}_i)[\nabla g(s)\cdot e_i+\frac{1}{2}e_i^{T}\nabla^{2}g(s)e_i]+N(s_i-s_{i+1})[\nabla g(s)\cdot (-e_i)+\frac{1}{2}e_i^{T}\nabla^{2}g(s)e_i] \nonumber \\
=&\sum_{i=1}^{b} \nabla g(s)\cdot[\lambda(s^{2}_{i-1}-s^{2}_i)-(s_i-s_{i+1})]Ne_i +\frac{1}{2}Ne_i^{T}\nabla^{2} g(s)e_i[\lambda(s^{2}_{i-1}-s^{2}_i)+(s_i-s_{i+1})] \nonumber \\
=&\nabla g(s)\cdot f(s)+\frac{1}{N}\sum_{i=1}^{b}\nabla^{2} g(s)_{ii} \tilde{f}_i(s). \nonumber
\end{align}
The first equality holds because $\nabla^{3} g(s)=0$ according to Lemma \ref{lem:Poi_sol}. \Halmos
\endproof

When the state $s$ is close to the equilibrium point $s^{*}$ in particular assuming $||s-s^{*}||^{2r}\le\frac{1}{N^{\epsilon}}$, we define $$x_i=s_i-s^{*}_i$$ and have the Taylor expansion of $\tilde{f}_i(s)$ at the equilibrium point $s^{*}$ 
\begin{align}
\tilde{f}_i(s)
=&\frac{1}{2}[\lambda(s^{2}_{i-1}-s^{2}_i)+(s_i-s_{i+1})] \nonumber \\
=&\frac{\lambda}{2}[(s^{*}_{i-1}+x_{i-1})^{2}-(s^{*}_i+x_i)^{2}] +\frac{1}{2}(s^{*}_i+x_i-s^{*}_{i+1}-x_{i+1}) \nonumber \\
=&\frac{\lambda}{2}[(s^{*}_{i-1})^{2}+2x_{i-1}s^{*}_{i-1}+x_{i-1}^{2}-(s^{*}_i)^{2}-2x_is^{*}_i-x_i^{2}]+\frac{1}{2}(s^{*}_i+x_i-s^{*}_{i+1}-x_{i+1}) \nonumber \\
=&\frac{\lambda}{2}[(s^{*}_{i-1})^{2}-(s^{*}_i)^{2}]+\frac{1}{2}(s^{*}_i-s^{*}_{i+1})+O(\frac{1}{N^{\frac{\epsilon}{2r}}}) \nonumber \\
=&\tilde{f}_i(s^{*})+O(\frac{1}{N^{\frac{\epsilon}{2r}}}) \label{eq: f_tilde},
\end{align} 
where the last equality holds because $||s-s^{*}||^{2r}\le\frac{1}{N^{\epsilon}}$ implies $|x_i|\le\frac{1}{N^{\frac{\epsilon}{2r}}}.$

Consider a state $s,$ which is close to the equilibrium point, i.e. $\|s-s^{*}\|^{2r}\le\frac{1}{N^{\epsilon}}.$ According to Stein's equation \eqref{eq: Poi_eq} and the previous lemma, we have 
\begin{align}
&Lg(s)-Gg(s)\nonumber\\
=&\nabla g(s)\cdot J(s^{*})(s-s^{*})-\nabla g(s)\cdot f(s) -\frac{1}{N}\sum_{i=1}^{b}\nabla^{2} g(s)_{ii} \tilde{f}_i(s)  \nonumber \\
=&\nabla g(s)\cdot \left(J(s^{*})(s-s^{*})-f(s)\right) -\frac{1}{N}\sum_{i=1}^{b}\nabla^{2} g(s)_{ii} \left(\tilde{f}_i(s^{*})+O\left(\frac{1}{N^{\frac{\epsilon}{2r}}}\right)\right) \nonumber \\
=&\nabla g(s)\cdot \left(J(s^{*})(s-s^{*})-f(s)\right)-\frac{1}{N}\sum_{i=1}^{b}\nabla^{2} g(s)_{ii}\tilde{f}_i(s^{*})-\frac{1}{N}\sum_{i=1}^{b}\nabla^{2} g(s)_{ii}O\left(\frac{1}{N^{\frac{\epsilon}{2r}}}\right). \nonumber
\end{align}
According to Lemma \ref{lem:Poi_sol}, we have 
\begin{align}
\nabla g(s)&=-2P(s-s^{*}), \nonumber \\
 \nabla^{2} g(s)&=-2P, \nonumber
\end{align}
which are the functions of $J(s^{*}).$ 

Since the original mean-field is a second-order system, we have  
\begin{align}
f(s)=&f(s^{*})+J(s^{*})(s-s^{*})+\frac{1}{2}<s-s^{*},\nabla^{2} f(s^{*})(s-s^{*})>, \nonumber
\end{align}
where $\nabla^{2} f(s^{*})$ is the Hessian of $f(s)$ at equilibrium point. For any $s\in\mathcal{S}$ and $i=1, \cdots, b$, the Hessian has the following form for $f_i(s)$
\[\nabla^{2} f_i(s)_{kj}=\frac{\partial^{2} f_i(s)}{\partial s_j\partial s_k}=\begin{cases}
-2\lambda, &\text{if } j=k=i, \\
2\lambda, &\text{if } j=k=i-1, \\
0, &\text{otherwise}. 
\end{cases}\]

Substituting it into the generator difference, we obtain
\begin{align}
&\mathbb E\left[Lg(S)-Gg(S)~\big\rvert~ S\in\mathcal B\right] \nonumber \\
=&\mathbb E\big[\underbrace{P(S-s^{*})\cdot<S-s^{*},\nabla^{2} f(s^{*})(S-s^{*})>}_{\text{Lemma \ref{lem: 1st_upp_bou}}}+\underbrace{\frac{2}{N}\sum_{i=1}^{b} P_{ii}\tilde{f}_i(s^{*})}_{\text{Lemmas \ref{lem: 2nd_low_bou} and \ref{lem: 2nd_upp_bou}}}
+\underbrace{\frac{2}{N}\sum_{i=1}^{b}P_{ii}O(\frac{1}{N^{\frac{\epsilon}{2r}}})}_{\text{Lemma \ref{lem: 3rd_upp_bou}}}~\big\rvert ~S\in\mathcal B\big].
\label{eq: gen_dif}
\end{align}
The above generator difference includes three terms. 

\begin{lemma}Given $||s-s^{*}||^{2r}\le\frac{1}{N^{\epsilon}}$, we have
\begin{align}
    ||P(s-s^{*})\cdot <s-s^{*},\nabla^{2} f(s^{*})(s-s^{*})>||= O\left(\frac{1}{N^{\frac{3\epsilon}{2r}-2\alpha-3\xi}}\right).
    \label{eq: 1st_upp_bou}
\end{align}
\label{lem: 1st_upp_bou}
\end{lemma}
\proof{Proof.}
Consider the 2-norm of the first term in \eqref{eq: gen_dif}. We have
\begin{align}
&\|P(s-s^{*})\cdot <s-s^{*},\nabla^{2} f(s^{*})(s-s^{*})>\| \nonumber \\
\leq &\|P(s-s^{*})\|\cdot \|<s-s^{*},\nabla^{2} f(s^{*})(s-s^{*})>\| \nonumber \\
\leq &\|P\|\cdot\|s-s^{*}\| \cdot\|<s-s^{*},\nabla^{2} f(s^{*})(s-s^{*})>\|  \nonumber \\
\le &2\sqrt{2}\lambda\|P\|\cdot \|s-s^{*}\|^{3},
\label{ine_1}
\end{align}
where the third inequality holds  because 
\begin{align}
\begin{split}
&\|<s-s^{*},\nabla^{2} f(s^{*})(s-s^{*})>\| \nonumber \\
=&\sqrt{\sum_{i=1}^{b}[(s-s^{*})\nabla^{2}f_i(s^{*})(s-s^{*})]^{2}} \nonumber \\
=&\sqrt{\sum_{i=1}^{b}\left(2\lambda[(s_{i-1}-s^{*}_{i-1})^{2}-(s_i-s^{*}_i)^{2}]\right)^{2}} \nonumber \\
=&2\lambda\sqrt{\sum_{i=1}^{b}[(s_{i-1}-s^{*}_{i-1})^{2}-(s_i-s^{*}_i)^{2}]^{2}} \nonumber \\
\le&2\lambda\sqrt{\sum_{i=1}^{b}(s_{i-1}-s^{*}_{i-1})^{4}+(s_i-s^{*}_i)^{4}} \nonumber \\
\le&2\sqrt{2}\lambda\sqrt{\sum_{i=1}^{b}(s_i-s^{*}_i)^{4}} \nonumber \\
\le&2\sqrt{2}\lambda\sqrt{[\sum_{i=1}^{b}(s_i-s^{*}_i)^{2}]^{2}} \nonumber \\
=&2\sqrt{2}\lambda\|s-s^{*}\|^{2}.\nonumber 
\end{split}
\end{align}
Furthermore, from Lemma \ref{lem: p_bound}, for a sufficiently large $N$, we have $\|P\| \le N^{2\alpha+3\xi}$. Since $||s-s^{*}||^{2r}\le\frac{1}{N^{\epsilon}}$, combining inequalities (\ref{ine_1}), we have
\begin{align}
    &\|P(s-s^{*})\cdot <s-s^{*},\nabla^{2} f(s^{*})(s-s^{*})>\| \nonumber \\
    \le & 2\sqrt{2}\lambda\times N^{2\alpha+3\xi}\times \frac{1}{N^{\frac{3\epsilon}{2r}}} =O\left(\frac{1}{N^{\frac{3\epsilon}{2r}-2\alpha-3\xi}}\right). \nonumber
\end{align} \Halmos
\endproof

\begin{lemma}
Given $||s-s^{*}||^{2r}\le\frac{1}{N^{\epsilon}}$, we have
\begin{align}
    \frac{2}{N}\sum_{i=1}^{b}P_{ii}(s)\tilde{f}_i(s^{*})&\ge\frac{\lambda\gamma}{9N^{1+\alpha}}.
     \label{eq: 2nd_low_bou}
\end{align}
\label{lem: 2nd_low_bou}
\end{lemma}
\proof{Proof.}
It is easy to check that for $i=1,\cdots,b$, $\tilde{f}_i(s^{*})\ge 0$. Also, from the fact that $P$ is a positive semi-definite, we have $P_{ii}\ge 0$. Therefore, for $i=1,\cdots,b$, we have
\begin{align}
    P_{ii}\tilde{f}_i(s^{*})\ge 0. \nonumber
\end{align}
Furthermore,  we also have
\begin{align}
\tilde{f}_i(s^{*})&=\frac{1}{2}[\lambda((s^{*}_{i-1})^{2}-(s^{*}_i)^{2})+(s^{*}_i-s^{*}_{i+1})] 
=\lambda[(s^{*}_{i-1})^{2}-(s^{*}_i)^{2}], \nonumber
\end{align}
where the second equality holds because $s^{*}$ is the equilibrium point. Thus, for $i=1$ by equation (\ref{eq: 1_b}), we have 
\begin{align}
\tilde{f}_1(s^{*})=\lambda[1-(s^{*}_1)^{2}]&\ge\lambda(1-\lambda^{2}) \geq \lambda(1-\lambda)=\frac{\lambda\gamma}{N^{\alpha}} \nonumber
\end{align}
which, together with Lemma \ref{lem: P_low_bound}, implies 
\begin{align}
\frac{2}{N}\sum_{i=1}^{b}P_{ii}\tilde{f}_i(s^{*})&\ge\frac{2}{N}P_{11}\tilde{f}_1(s^{*})\ge\frac{\lambda\gamma}{9N^{1+\alpha}}. \nonumber
\end{align} \Halmos
\endproof

\begin{lemma}Given $||s-s^{*}||^{2r}\le\frac{1}{N^{\epsilon}}$, we have
\begin{align}
     \frac{2}{N}\sum_{i=1}^{b}P_{ii}\tilde{f}_i(s^{*})=O(\frac{1}{N^{1-2\alpha-4\xi}}).
      \label{eq: 2nd_upp_bou}
\end{align}
\label{lem: 2nd_upp_bou}
\end{lemma}
\proof{Proof.}
It is easy to check that $\tilde{f}_i(s^{*})\le 1$ for $i=1,\cdots,b$. Recall that $P_{ii}\le \|P\|\le N^{2\alpha+3\xi}$. Therefore, we have
\begin{align}
    \frac{2}{N}\sum_{i=1}^{b}P_{ii}\tilde{f}_i(s^{*}) 
    =\frac{b}{N} O(N^{2\alpha+3\xi})=O(\frac{1}{N^{1-2\alpha-4\xi}}). \nonumber
\end{align} \Halmos
\endproof

\begin{lemma} Given $||s-s^{*}||^{2r}\le\frac{1}{N^{\epsilon}}$, for arbitrarily small $\xi>0$, there is a sufficiently large $N$ that the following holds
\begin{align}
    \left\|\frac{2}{N}\sum_{i=1}^{b}P_{ii}O(\frac{1}{N^{\frac{\epsilon}{2r}}})\right\| = O\left(\frac{1}{N^{1+\frac{\epsilon}{2r}-2\alpha-4\xi}}\right).
     \label{eq: 3rd_upp_bou}
\end{align}
\label{lem: 3rd_upp_bou}
\end{lemma}
\proof{Proof.}
Recall that $|P_{ii}|\le N^{2\alpha+3\xi}$ for $i=1,\cdots,b$. Thus, we have
\begin{align}
    \left\|\frac{2}{N}\sum_{i=1}^{b}P_{ii}O(\frac{1}{N^{\frac{\epsilon}{2r}}})\right\| 
    \le\frac{2b}{N}N^{2\alpha+3\xi}\cdot O(\frac{1}{N^{\frac{\epsilon}{2r}}}) 
    =O(\frac{1}{N^{1+\frac{\epsilon}{2r}-2\alpha-4\xi}}). \nonumber
\end{align} \Halmos
\endproof

Based on these lemmas, we  are now able to characterize the generator difference when  state $S$ is close to $s^*.$
\begin{lemma}
For $0<\alpha<\frac{1}{18}$ and arbitrarily small $\xi>0$, there is a sufficiently large $N$ that the following holds
\begin{align}
    & \mathbb E\left[Lg(S)-Gg(S)\left| S\in\mathcal B\right.\right]
    =\frac{2}{N}\sum_{i=1}^{b} P_{ii}\tilde{f}_i(s^{*})+o\left(\frac{1}{N^{1+\alpha}}\right),
\end{align}
with the following choice of parameters
\begin{align}
\frac{3(1+\alpha+\xi)}{1-18\alpha-27\xi}&< r \label{ineq: r}, \\
\frac{2r(1+3\alpha+3\xi)}{3}&<\epsilon<r(1-4\alpha-7\xi)-1-\alpha-\xi
\label{ineq: epsilon}. 
\end{align}
\label{lem: dom_ter}
\end{lemma}
\proof{Proof.}
Under the conditions of the lemma, it is easy to check that the upper bound of the first and third terms in equation (\ref{eq: gen_dif}) are order-wise smaller than the lower bounds of the second term, i.e.
\begin{align}
\frac{3\epsilon}{2r}-2\alpha-3\xi>1+3\alpha+3\xi-2\alpha-3\xi=1+\alpha \nonumber
\end{align}
and 
\begin{align}
1+\frac{\epsilon}{2r}-2\alpha-4\xi&>1+\frac{1}{3}+\alpha+\xi-2\alpha-4\xi \nonumber \\
&>(1+\alpha)+(\frac{2}{9}-3\xi), \nonumber
\end{align}
where the last inequality is by the fact $0<\alpha<\frac{1}{18}$. Therefore, the lemma holds. \Halmos
\endproof

We also remark that there exist parameters that satisfy the conditions in the lemma because the right-hand side of $\epsilon$ in \eqref{ineq: epsilon} is larger than the left-hand side given that the $r$  satisfies \eqref{ineq: r}, where $r$ has to be large enough.
For example, when $\alpha=0.05$, $r$ needs to at least 32 and $\epsilon$ can be $24.54$, and it's easy to check that we can find a small enough $\xi$.

To present the main theorem, we first provide the following lemma that ensure the invertibility that is required in the expression calculable error bound.
\begin{lemma}
    The matrix, $J^{\top}(s^{*})\otimes I+I\otimes J^{\top}(s^{*})$, is invertible.
    \label{lem: Kro_inv}
\end{lemma}
\proof{Proof.}
Since $J(s^*)$ is Hurwitz, any eigenvalue $\mu_i$ has negative real parts, so is the eigenvalues of $J^{\top}(s^*)$. Furthermore, the eigenvalues of $J^{\top}(s^{*})\otimes I$ are the product of eigenvalues of $J^{\top}(s^*)$ and $I$. In other words, all $\mu_i = \mu_i\times 1$ are eigenvalues. 

Let $A=J^{\top}(s^{*})\otimes I+I\otimes J^{\top}(s^{*})$. For any eigen vectors of $v_i$ and $v_j$ of $J^{\top}(s^{*})$ that has
\begin{align}
J^{\top}(s^{*}) v_i &= \mu_i v_i \nonumber \\
J^{\top}(s^{*}) v_j &= \mu_j v_j \nonumber,
\end{align}
then we have
\begin{align}
A(v_i\otimes v_j) &= (J^{\top}(s^{*})\otimes I)(v_i\otimes v_j)+(I\otimes J^{\top}(s^{*}))(v_i\otimes v_j) \nonumber \\
&= (J^{\top}(s^{*})v_i)\otimes v_j+v_i\otimes J^{\top}(s^{*})v_j \nonumber \\
&= \mu_i v_i\otimes v_j + v_i\otimes \mu_j v_j \nonumber \\
&= (\mu_i+\mu_j) v_i\otimes v_j. \nonumber
\end{align}
In other words, the eigenvalues of $J^{\top}(s^{*})\otimes I+I\otimes J^{\top}(s^{*})$ are simply the sum of respetive eigenvalues, i.e. $\mu_i + \mu_j$. Since $J(s^*)$ is Hurwitz, $\mu_i + \mu_j$ also has negative real parts and not zero. Therefore, $J^{\top}(s^{*})\otimes I+I\otimes J^{\top}(s^{*})$ is invertible.
\Halmos

\subsection{Proof of Theorem \ref{thm:convergence}}
The proof consists of 2 steps. First, we establish the dominant term in all state space; Then, based on the observation of the dominant term being the trace of product of two matrices, we provide the calculable error bound.

\textbf{Step 1}:
We again choose parameters that satisfy the following conditions: 
\begin{align}
\frac{3(1+\alpha)}{1-18\alpha-27\xi}&< r \nonumber \\
\frac{2r(1+3\alpha+3\xi)}{3}&<\epsilon<r(1-4\alpha-7\xi)-1-\alpha-\xi \nonumber
\end{align}
and $\xi>0$ is arbitrarily small. Then, for a sufficiently large $N$, the mean square distance is
\begin{align}
&\mathbb E[||S-s^{*}||^{2}] \nonumber \\
=&\mathbb E\left[||S-s^{*}||^{2}\big\rvert S\notin\mathcal B\right]\mathbb{P}\left(S\notin\mathcal B\right)+\mathbb E\left[||S-s^{*}||^{2}\big\rvert S\in\mathcal B\right]\mathbb{P}\left(S\in\mathcal B\right)  \nonumber \\
=&O(\log N)\times O(\frac{1}{N^{r(1-4\alpha-7\xi)-\epsilon}})+
\left(-\frac{1}{N}\sum_{i=1}^{b}\nabla^{2} g(s)_{ii} \tilde{f}_i(s^{*}) 
+o(\frac{1}{N^{1+\alpha}})\right)
\times \left(1-O(\frac{1}{N^{r(1-4\alpha-7\xi)-\epsilon}})\right) \nonumber \\
=&O\left(\frac{1}{N^{r(1-4\alpha-7\xi)-\epsilon-\xi}}\right)-\frac{1}{N}\sum_{i=1}^{b}\nabla^{2} g(s)_{ii} \tilde{f}_i(s^{*}) +O\left(\frac{1}{N^{1-2\alpha-3\xi}}\right)\times O\left(\frac{1}{N^{r(1-4\alpha-7\xi)-\epsilon}}\right)+o\left(\frac{1}{N^{1+\alpha}}\right)\nonumber \\
=&\frac{2}{N}\sum_{i=1}^{b} P_{ii}\tilde{f}_i(s^{*})+o\left(\frac{1}{N^{1+\alpha}}\right), \nonumber
\end{align}
where the second equality holds because $||s-s^{*}||^{2}\le b=O(\log N).$
Note that with the choice of parameters $r, \epsilon$ and $0<\alpha<\frac{1}{18},$ the lower bound of the term $\frac{2}{N}\sum_{i=1}^{b} P_{ii}\tilde{f}_i(s^{*})+o(\frac{1}{N^{1+\alpha}})$ is $O(\frac{1}{N^{1+\alpha}})$, while the other terms are strictly upper bounded by this order for sufficiently large $N$.

\textbf{Step 2}:
Recall the Lyapunov equation 
\begin{align}
PJ(s^{*})+J^{\top}(s^{*})P = -I \label{eq: Lyapunov_eq}
\end{align}
and the goal is to obtain $\text{tr}(P\tilde{F})$, where $ \tilde{F} = \text{diag}(\tilde{f}_1(s^{*}),\cdots, \tilde{f}_b(s^{*}))$. We observe that $\text{tr}(P\tilde{F}) = \text{vec}(\tilde{F})^{\top}\text{vec}(P)$.
Vectorizing the Lyapunov equation \eqref{eq: Lyapunov_eq}, for the left-hand side (LHS), we have
\begin{align}
\text{LHS} &= \text{vec}(PJ(s^{*}))+ \text{vec}(J^{\top}(s^{*})P) \nonumber \\
&= \text{vec}(IPJ(s^{*}))+ \text{vec}(J^{\top}(s^{*})PI) \nonumber \\
&= (J^{\top}(s^{*})\otimes I)\text{vec}(P) + (I\otimes J^{\top}(s^{*}))\text{vec}(P) \nonumber \\
&= (J^{\top}(s^{*})\otimes I+I\otimes J^{\top}(s^{*}))\text{vec}(P) \nonumber 
\end{align}
where in the third equality we used the fact $\text{vec}(ABC)=(C^{\top}\otimes A)\text{vec}(B)$. Similarly, the vectorization of the right-hand side of Lyapunov equation \eqref{eq: Lyapunov_eq}, we simpley have $-\text{vec}(I)$. As a result, the vectorized Lyapunov equation yields
\begin{align}
(J^{\top}(s^{*})\otimes I+I\otimes J^{\top}(s^{*}))\text{vec}(P) = -\text{vec}(I). \nonumber
\end{align}
By invertibility in Lemma \ref{lem: Kro_inv}, we have
$$\text{vec}(P) = -(J^{\top}(s^{*})\otimes I+I\otimes J^{\top}(s^{*}))^{-1}\text{vec}(I).$$ Finally, we have
\begin{align}
\sum_{i=1}^{b} P_{ii}\tilde{f}_i(s^{*}) = \text{tr}(P\tilde{F}) = -\text{vec}(\tilde{F})^{\top}(J^{\top}(s^{*})\otimes I+ I\otimes J^{\top}(s^{*}))^{-1} \text{vec}(I). \nonumber
\end{align}
\Halmos

\subsection{Proof of the Remark \ref{thm: order-wise}}
We choose following parameter choices 
\begin{align}
&\frac{3(1-2\alpha-2\xi)}{1-12\alpha-21\xi}\le r \label{eq: r_ord_wis} \\
&\frac{2r}{3}\le\epsilon\le r(1-4\alpha-7\xi)-1-\alpha-\xi \nonumber
\end{align}
and $\xi>0$ is arbitrarily small. 
For $\alpha\in(0,1/12)$, similar to the proof of Theorem \ref{thm:convergence}, we show
that the term in equation \eqref{eq: 2nd_upp_bou} in Lemma \ref{lem: 2nd_upp_bou} is the dominant term. It is easy to check for $s\in\mathcal{B}$ that 
\begin{align}
    & \mathbb E\left[Lg(S)-Gg(S)\left| S\in\mathcal B\right.\right]
    =O(\frac{1}{N^{1-2\alpha-3\xi}}). 
    \label{eq: upp_bou_ord_wis}
\end{align}
Combined with the probability of $s\notin \mathcal{B}$ and equation \eqref{eq: upp_bou_ord_wis}, for sufficiently large $N$, we have the mean square error as
\begin{align}
&\mathbb E[||S-s^{*}||^{2}] \nonumber \\
=&\mathbb E\left[||S-s^{*}||^{2}\big\rvert S\notin\mathcal B\right]\mathbb{P}\left(S\notin\mathcal B\right)+\mathbb E\left[||S-s^{*}||^{2}\big\rvert S\in\mathcal B\right]\mathbb{P}\left(S\in\mathcal B\right)  \nonumber \\
=&O(\log N)\times O(\frac{1}{N^{r(1-4\alpha-7\xi)-\epsilon}})+O(\frac{1}{N^{1-2\alpha-3\xi}})
\times \left(1-O(\frac{1}{N^{r(1-4\alpha-7\xi)-\epsilon}})\right) \nonumber \\
=&O\left(\frac{1}{N^{r(1-4\alpha-7\xi)-\epsilon-\xi}}\right)+O(\frac{1}{N^{1-2\alpha-3\xi}}) \nonumber \\
=&O(\frac{1}{N^{1-2\alpha-3\xi}}) \nonumber
\end{align}
where the last equality is the result of the parameter choices. We note that the requirement for $\alpha<\frac{1}{12}$ results from the fact that the denominator of LHS of inequality \eqref{eq: r_ord_wis} needs to be positive.

As a result for $\alpha\in(0,\frac{1}{18})$ where the calculable asymptotic tight bound holds, the order-wise convergence result above also holds.

%Together with Lemma \ref{lem: pre_con}, we conclude the proof for all $\alpha\in(0,0.25)$, hence the Theorem \ref{thm: order-wise} holds.

\section{Conclusion}
In this paper, we established a calculable bound on the mean-square errors of the power-of-two-choices mean-field model in heavy-traffic. Our approach combined SSC and Stein's method with a linearized mean-field models, and characterized the dominant term of the mean square error. Our simulation results confirmed the theoretical bound and showed that the bounds are valid even for small size systems such as when $N=100.$ This recipe of combining SSC and Stein's method for linearized mean-field model can be applied to other mean-field models beyond the power-of-two-choices load balancing algorithm.

\section*{Acknowledgements} 
The authors are very grateful to Nicolas Gast for his valuable comments. This work
was supported in part by NSF ECCS 1739344, CNS 2002608 and CNS 2001687.

%%%%%%%%%%%%%%%%%%%%%%%%%%%%%%%%%%%%%%%%%%%%%%%%%%%%%%%%%%%%%%%%%%%%%%%%%%%%
%% Appendix
%%%%%%%%%%%%%%%%%%%%%%%%%%%%%%%%%%%%%%%%%%%%%%%%%%%%%%%%%%%%%%%%%%%%%%%%%%%%

\appendix

\section{Proof of Lemma \ref{lem: pre_con}}\label{sec: pre_con}
The proof and analysis follow from the heavy-traffic infinite buffer size case in \citep{Yin_17}. We remark that the differences between the mean-field model here and the truncated mean-field model in Lemma 4.2 of \citep{Yin_17} are the dimension and the last equation. Specifically, we used dimension $b=O(\log N)$ and the truncated system considers $N^{\alpha+\xi}$ dimensional system, where $\xi>0$ is arbitrarily small; for the dynamical equation in the last dimension, we don't have an added term and, in \citep{Yin_17}, a term is added so that the solution of the mean-field model can be written in a closed form. Due to the buffer size assumption of $b=O(\log N)$, we can allow $\alpha$ to be in the range of $(0,0.25)$. We omit the proof in here.

\subsection{Bounds on $e(t)$}
Following the analysis in \citep{Yin_16}, we consider the following collection of dynamical systems:
\begin{align}
    \dot{e}(t)&=f(x(t,z))-f(x(t,y))-\frac{1}{N}\frac{\partial f}{\partial x}(x(t,y))x^{(1)}(t) \nonumber \\
    \dot{x}^{(1)}(t)&=\frac{\partial f}{\partial x}(x(t,y))x^{(1)}(t) \nonumber \\
    \dot{x}(t,y)&=f(x(t,y)) \nonumber \\
    \dot{x}(t,z)&=f(x(t,z)) \nonumber
\end{align}
with initial conditions $e(0)=0, x^{(1)}(0)=N(z-y), x(0,z)=z$ and $x(0,y)=y$, where $\frac{\partial f}{\partial x}$ denotes the Jacobian matrix. It is proven that for Po2, it satisfies Conditions C1 -- C7 in Lemme 2.2 \citep{Yin_17}. We derive the following results as a result of Lemma 2.2 in \citep{Yin_17} for Po2.

To bound higher moment bounds, we first establish bounds on $|e(t)|$ for any $t\ge 0$ and $\int_0^{\infty}|e(t)|dt$.
\begin{lemma}
Given an arbitrarily small $\xi>0$, there's a sufficiently large $N$, the following bounds on term $e(t)$ and its integral hold true
\begin{align}
||e(t)||&\le|e(t)|\le \frac{1}{N^{2-2\alpha-4\xi}}\qquad \forall t\ge 0 \label{eq: e_bou}\\
\int_0^{\infty}|e(t)|dt&\le \frac{1}{N^{2-4\alpha-5\xi}}.
\end{align}
\end{lemma}
\proof{Proof.}
According to C4 and C6 in Lemma 2.2 in \citep{Yin_17}, and the fact $e(0)=0$, we have that for $t\le\tilde{t}_{d,z}$,
\begin{align}
|e(t)|&\le\frac{c_e}{N^{2}}t \nonumber \\
&\le\frac{c_e}{N^{2}}t_{d,z} \nonumber \\
&\le\frac{c_e}{N^{2}}\frac{1}{\delta_0}\max\{0,\log24|x(0)|\} \nonumber \\
&\le\frac{c_e}{N^{2}}\cdot \frac{12N^{2\alpha+2\xi}}{\gamma}\cdot 24|x(0)| \nonumber \\
&\le\frac{288c_e}{\gamma N^{2-2\alpha-2\xi}} b\nonumber \\
&\le\frac{1}{N^{2-2\alpha-3\xi}}. \nonumber
\end{align}
The last inequality is because $b=O(\log N)$.

For $t\ge \tilde{t}_{d,z}$, from C5 and comparison principle, we obtain
\begin{align}
|e(t)|&\le\frac{1}{c_{le}}V_e(e(t)) \nonumber \\
&\le V_e(e(\tilde{t}_{d,z}))e^{-\delta_e(t-\tilde{t}_{d,z})}+\frac{c}{N^{2}}e^{-\delta_e(t-\tilde{t}_{d,z})}\frac{1}{\alpha\delta_1-\delta_e}(1-\exp(-(\alpha_e\delta_1-\delta_e)(t-\tilde{t}_{d,z}))) \nonumber
\end{align}
where $\tilde{t}_{d,z}=\frac{1}{\delta_0}\max\{0,\log24|x(0)|\}$, $\delta_1=\delta_e=\tilde{\delta}$, $\alpha_e=2$ and $c=c_{er}(\frac{c_{u1}c_1}{c_{l1}})^{\alpha_e}$. Hence
\begin{align}
|e(t)|&\le V_e(e(\tilde{t}_{d,z}))+\frac{c}{N^{2}}\frac{1}{\tilde{\delta}} \nonumber \\
&\le c_{ue}|e(\tilde{t}_{d,z})|+\frac{c}{N^{2}}\frac{1}{\tilde{\delta}} \nonumber \\
&\le c_{ue}\frac{1}{N^{2-2\alpha-3\xi}}+\frac{c}{N^{2}}\frac{3N^{2\alpha+2\xi}}{\gamma} \nonumber \\
&\le \frac{1}{N^{2-2\alpha-4\xi}} \nonumber
\end{align}
where the last inequality holds for sufficiently large $N$. Furthermore, we have 
\begin{align}
    ||e(t)||\le |e(t)|. \nonumber
\end{align}

For the integral term, we have
\begin{align}
\int_{0}^{\infty}|e(t)|dt&\le \frac{\kappa}{\gamma^{2}}(144\frac{\max^{2}\{0,\ln24|x(0)|\}}{N^{2-4\alpha-4\xi}}+36\frac{\max\{0,\ln24|x(0)|\}}{N^{2-4\alpha-4\xi}}+36\frac{1}{N^{2-4\alpha-4\xi}}) \nonumber \\
&\le\frac{1}{N^{2-4\alpha-5\xi}} \nonumber 
\end{align}
where the last inequality holds for sufficiently large $N$. \Halmos
\endproof

\section{Proof of Lemma \ref{lem: hig_mom}}
\proof{Proof.}
In this proof, we use mathematical induction. When $r=1$, it holds because of Lemma \ref{lem: pre_con}. Assuming that \eqref{eq: hig_mom_bou} holds for $r-1$, we show that it also holds for $r$.
We use Stein's method to bound the distance. However, we consider a distance function that is $||s-s^{*}||^{2r}$, where $r\in\mathbb{N}$ is an integer.
Let $x=s-s^{*}$ and $X=S-s^{*}$, so the goal is to bound $\mathbb{E}[||X||^{2r}]$.

Consider a function $g_r: \mathcal{X}\to R$ such that it is the solution to the following Stein's equation,
\begin{align}
\nabla g_r(x)\cdot \dot{x}=\nabla g_r(x)\cdot f(x)=||x||^{2r}. \nonumber
\end{align}
Then, the solution has the following form 
\begin{align}
g_r(x)&=-\int_{0}^{\infty}||x(t,x)||^{2r}dt \nonumber \\
&=-\int_{0}^{\infty}[\sum_{i=1}^{b}x^{2}_i(t,x)]^{r}dt. \nonumber
\end{align}
We have for the stationary distribution of $X$
\begin{align}
\mathbb{E}[Gg_r(X)]=\mathbb{E}[\sum_{y\neq X}R_{X,y}(X)(g_r(y)-g_r(X))]=0. \nonumber
\end{align}
Recall that $q_{X,y}=\frac{1}{N}R_{X,y}$ for all $X, y\in\mathcal{X}$.
Then, we will have 
\begin{align}
&\mathbb{E}[||X||^{2r}] \nonumber \\
=&\mathbb{E}[\nabla g_r(X)\cdot f(X)-N\sum_{y\neq X}q_{X,y}(g_r(y)-g_r(X))] \nonumber \\
=&\mathbb{E}[\nabla g_r(X)\cdot f(X)-\nabla g_r(X)\cdot \sum_{y\neq X}q_{X,y}N(y-X)
+\nabla g_r(X)\cdot \sum_{y\neq X}q_{X,y}N(y-X)-N\sum_{y\neq X}q_{X,y}(g_r(y)-g_r(X))] \nonumber \\
=&\mathbb{E}[\nabla g_r(X)\cdot\left(f(X)-\sum_{y\neq X}q_{X,y}N(y-X)\right)
-\sum_{y\neq X}q_{X,y}N(g_r(y)-g_r(X)-\nabla g_r(X)\cdot(y-X))] \nonumber \\
=&\mathbb{E}[-\sum_{y\neq X}q_{X,y}N(g_r(y)-g_r(X)-\nabla g_r(X)\cdot(y-X))]. \nonumber 
\end{align}

Next, we focus on the following term
\begin{align}
&-(g_r(y)-g_r(x)-\nabla g_r(x)\cdot(y-x)) \nonumber \\
=&\int_{0}^{\infty}[\sum_{i=1}^{b}x^{2}_i(t,y)]^{r}dt-\int_{0}^{\infty}[\sum_{i=1}^{b}x^{2}_i(t,x)]^{r}dt-2r(y-x)\cdot\int_{0}^{\infty}[\sum_{i=1}^{b}x^{2}_i(t,x)]^{r-1} \cdot\sum_{i=1}^{b}x_i(t,x)\nabla x_i(t,x)dt\nonumber\\
=&\int_{0}^{\infty}\left([\sum_{i=1}^{b}x^{2}_i(t,y)]^{r}-[\sum_{i=1}^{b}x^{2}_i(t,x)]^{r}-2r(y-x)\cdot[\sum_{i=1}^{b}x^{2}_i(t,s)]^{r-1} \cdot\sum_{i=1}^{b}x_i(t,x)\nabla x_i(t,x)\right)dt.  
\label{eq: gr_tay}
\end{align}
Define $e(t)$, the same as in Ying (2016),
\begin{align}
e_i(t)=x_i(t,y)-x_i(t,x)-\nabla x_i(t,x)\cdot(y-x) \nonumber 
\end{align}
i.e.,
\begin{align}
x_i(t,y)=e_i(t)+x_i(t,x)+\nabla x_i(t,x)\cdot (y-x) \nonumber
\end{align}
where $x(t,x)$ denotes the trajectory of the mean-field dynamical system with $x$ as the initial condition and $\nabla x(t,x)$ refers to differentiating with respect to the initial condition $x$. 
So 
\begin{align}
&x^{2}_i(t,y) \nonumber \\
=&(e_i(t)+x_i(t,x)+\nabla x_i(t,x)\cdot (y-x))^{2} \nonumber \\
=&x^{2}_i(t,x)+e_i^{2}(t)+[\nabla x_i(t,x)\cdot (y-x)]^{2}+2e_i(t)x_i(t,x)
+2x_i(t,x)\nabla x_i(t,x)\cdot (y-x)+2e_i(t)\nabla x_i(t,x)\cdot(y-x).\nonumber
\end{align}

By summing over all $i$ and raising to the $r$th order, we have 
\begin{align}
&[\sum_{i=1}^{b}x^{2}_i(t,y)]^{r} \nonumber \\
=&\sum_{\sum_{k=1}^{6}r_k=r, r_k\ge0}{r \choose r_1}{r-r_1 \choose r_2}{r-r_1-r_2 \choose r_3}{r-r_1-r_2-r_3 \choose r_4} {r-r_1-r_2-r_3-r_4 \choose r_5} \nonumber \\
&[\sum_{i=1}^{b}x^{2}_i(t,x)]^{r_1}[\sum_{i=1}^{b}e_i^{2}(t)]^{r_2}[\sum_{i=1}^{b}[\nabla x_i(t,x)\cdot (y-x)]^{2}]^{r_3}[2\sum_{i=1}^{b}e_i(t)x_i(t,x)]^{r_4} \nonumber \\
&[2\sum_{i=1}^{b}x_i(t,x)\nabla x_i(t,x)\cdot (y-x)]^{r_5}[2\sum_{i=1}^{b}e_i(t)\nabla x_i(t,x)\cdot(y-x)]^{r_6}. \nonumber 
\end{align}
When $r_1=r$ and $r_i=0$ for $i=2,\cdots,6$, the summand is $[\sum_{i=1}^{b}x^{2}_i(t,x)]^{r}$ and when $r_1=r-1, r_5=1$ and $r_i=0$ for $i=2,3,4,6$, the summand is 
$2r[\sum_{i=1}^{b}x^{2}_i(t,x)]^{r-1}\sum_{i=1}^{b}x_i(t,x)\nabla x_i(t,x)\cdot (y-x)$. Let $\Sigma$ be the collection of combination of all $\{r_i\}_{i=1,\cdots,6}$ that excludes above two cases, i.e.
$$\Sigma=\{r_i,i=1,\cdots,6|\sum_{i=1}^{6}=r\}\setminus \{\{r_1=r,r_i=0, \text{for}\quad i=2,\cdots,6\},\{r_1=r-1,r_5=1,r_i=0, \text{for}\quad i=2,3,4,6\}\}.$$

From \eqref{eq: e_bou}, we have for any $i=1,\cdots,b$  
$$|e_i(t)|\le|e(t)|\le\frac{1}{N^{2-2\alpha-4\xi}},$$
and the state transition condition
$$||x-y||=\frac{1}{N}.$$
From Ying (2016), we have
\begin{align}
x^{(1)}(t)=\nabla x(t,x)\cdot N(y-x), \nonumber 
\end{align}
equivalent to, $$\nabla x(t,x)\cdot(y-x)=\frac{1}{N}x^{(1)}(t).$$
Therefore, $|\nabla x(t,x)\cdot(y-x)|=\frac{1}{N}|x^{(1)}(t)|\le\frac{1}{N}|x^{(1)}(0)|=\frac{1}{N}$, by Lemma C.6 in \citep{Yin_17}. Also
\begin{align}
||\nabla x(t,x)\cdot(y-x)||^{2}&=\sum_{i=1}^{b}[\nabla x_i(t,x)\cdot(y-x)]^{2} \nonumber \\
&=\sum_{i=1}^{b}|\frac{1}{N}x^{(1)}_i(t)|^{2} \nonumber \\
&=\frac{1}{N^{2}}\sum_{i=1}^{b}|x^{(1)}_i(t)|^{2} \nonumber \\
&\le \frac{1}{N^{2}}\sum_{i=1}^{b}|x^{(1)}_i(t)| \nonumber \\
&\le \frac{1}{N^{2}}|x^{(1)}(t)|\le\frac{1}{N^{2}}|x^{(1)}(0)|=\frac{1}{N^{2}}. \nonumber
\end{align}
The last inequality is because $|x^{(1)}(t)|\le|x^{(1)}(0)|=1$. So, for any $i\in\{1,\cdots,b\}$ ,we have 
$$|\nabla x_i(t,x)\cdot(y-x)|\le||\nabla x(t,x)\cdot(y-x)||\le\frac{1}{N}.$$
Then, for the terms inside integral in \eqref{eq: gr_tay}, we have
\begin{align}
&[\sum_{i=1}^{b}x^{2}_i(t,y)]^{r}-[\sum_{i=1}^{b}x^{2}_i(t,x)]^{r}-2r(y-x)\cdot[\sum_{i=1}^{b}x^{2}_i(t,x)]^{r-1}\cdot\sum_{i=1}^{b}x_i(t,x)\nabla x_i(t,x) \nonumber \\
=&\sum_{\Sigma}{r \choose r_1}{r-r_1 \choose r_2}{r-r_1-r_2 \choose r_3}{r-r_1-r_2-r_3 \choose r_4}{r-r_1-r_2-r_3-r_4 \choose r_5} \nonumber \\
&[\sum_{i=1}^{b}x^{2}_i(t,x)]^{r_1}[\sum_{i=1}^{b}e_i^{2}(t)]^{r_2}[\sum_{i=1}^{b}[\nabla x_i(t,x)\cdot (y-x)]^{2}]^{r_3}[2\sum_{i=1}^{b}e_i(t)x_i(t,x)]^{r_4} \nonumber \\
&[2\sum_{i=1}^{b}x_i(t,x)\nabla x_i(t,x)\cdot (y-x)]^{r_5} 
[2\sum_{i=1}^{b}e_i(t) \nabla x_i(t,x)\cdot(y-x)]^{r_6} \nonumber \\
\le&\sum_{\Sigma}{r \choose r_1}{r-r_1 \choose r_2}{r-r_1-r_2 \choose r_3}{r-r_1-r_2-r_3 \choose r_4}{r-r_1-r_2-r_3-r_4 \choose r_5} \nonumber \\
&[\sum_{i=1}^{b}x^{2}_i(t,x)]^{r_1}[\sum_{i=1}^{b}e_i^{2}(t)]^{r_2}[\sum_{i=1}^{b}[\nabla x_i(t,x)\cdot (y-x)]^{2}]^{r_3}[2\sum_{i=1}^{b}|e_i(t)|\cdot|x_i(t,x)|]^{r_4} \nonumber \\
&[2\sum_{i=1}^{b}|x_i(t,x)|\cdot|\nabla x_i(t,x)\cdot (y-x)|]^{r_5} 
[2\sum_{i=1}^{b}|e_i(t)|\cdot |\nabla x_i(t,x)\cdot (y-x)|]^{r_6} \nonumber \\
=&\sum_{\Sigma}{r \choose r_1}{r-r_1 \choose r_2}{r-r_1-r_2 \choose r_3}{r-r_1-r_2-r_3 \choose r_4}{r-r_1-r_2-r_3-r_4 \choose r_5} \nonumber \\
&[\sum_{i=1}^{b}x^{2}_i(t,x)]^{r_1}||e(t)||^{2r_2}||\nabla x(t,x)\cdot (y-x)||^{2r_3}[2\sum_{i=1}^{b}|e_i(t)|\cdot|x_i(t,x)|]^{r_4} \nonumber \\
&[2\sum_{i=1}^{b}|x_i(t,x)|\cdot|\nabla x_i(t,x)\cdot (y-x)|]^{r_5} 
[2\sum_{i=1}^{b}|e_i(t)|\cdot |\nabla x_i(t,x)\cdot (y-x)|]^{r_6} \nonumber \\
\le&\sum_{\Sigma}{r \choose r_1}{r-r_1 \choose r_2}{r-r_1-r_2 \choose r_3}{r-r_1-r_2-r_3 \choose r_4}{r-r_1-r_2-r_3-r_4 \choose r_5} \nonumber \\
&[\sum_{i=1}^{b}x^{2}_i(t,x)]^{r_1}\frac{1}{N^{2(2-2\alpha-4\xi)r_2}}\frac{1}{N^{2r_3}}\frac{2^{r_4}}{N^{(2-2\alpha-4\xi)r_4}}[\sum_{i=1}^{b}|x_i(t,x)|]^{r_4}
\frac{2^{r_5}}{N^{r_5}}[\sum_{i=1}^{b}|x_i(t,x)|]^{r_5}\frac{2^{r_6}b^{r_6}}{N^{(3-2\alpha-4\xi)r_6}} \nonumber \\
\le&\sum_{\Sigma}{r \choose r_1}{r-r_1 \choose r_2}{r-r_1-r_2 \choose r_3}{r-r_1-r_2-r_3 \choose r_4}{r-r_1-r_2-r_3-r_4 \choose r_5} \nonumber \\
&\frac{2^{r_4+r_5+r_6}b^{r_6}}{N^{2(2-2\alpha-4\xi)r_2+2r_3+(2-2\alpha-4\xi)r_4+r_5+(3-2\alpha-4\xi)r_6}}[\sum_{i=1}^{b}x^{2}_i(t,x)]^{r_1}
[\sum_{i=1}^{b}|x_i(t,x)|]^{r_4+r_5}. \nonumber
\end{align}

Then, by substituting into the equation \eqref{eq: gr_tay}, we have
\begin{align}
&-(g_r(y)-g_r(x)-\nabla g_r(x)\cdot(y-x)) \nonumber \\
=&\int_{0}^{\infty}\left([\sum_{i=1}^{b}x^{2}_i(t,y)]^{r}-[\sum_{i=1}^{b}x^{2}_i(t,x)]^{r}-2r(y-x)\cdot[\sum_{i=1}^{b}x^{2}_i(t,s)]^{r-1} \cdot\sum_{i=1}^{b}x_i(t,x)\nabla x_i(t,x)\right)dt \nonumber \\
\le &\sum_{\Sigma}{r \choose r_1}{r-r_1 \choose r_2}{r-r_1-r_2 \choose r_3}{r-r_1-r_2-r_3 \choose r_4}{r-r_1-r_2-r_3-r_4 \choose r_5} \nonumber \\
&\frac{2^{r_4+r_5+r_6}b^{r_6}}{N^{2(2-2\alpha-4\xi)r_2+2r_3+(2-2\alpha-4\xi)r_4+r_5+(3-2\alpha-4\xi)r_6}}\int_{0}^{\infty}[\sum_{i=1}^{b}x^{2}_i(t,x)]^{r_1}[\sum_{i=1}^{b}|x_i(t,x)|]^{r_4+r_5}dt \nonumber \\
\le &\sum_{\Sigma}{r \choose r_1}{r-r_1 \choose r_2}{r-r_1-r_2 \choose r_3}{r-r_1-r_2-r_3 \choose r_4}{r-r_1-r_2-r_3-r_4 \choose r_5} \nonumber \\
&\frac{2^{r_4+r_5+r_6}b^{r_6}}{N^{2(2-2\alpha-4\xi)r_2+2r_3+(2-2\alpha-4\xi)r_4+r_5+(3-2\alpha-4\xi)r_6}}\int_{0}^{\infty}(3|x|e^{-\delta_0 t})^{2r_1}(3|x|e^{-\delta_0 t})^{r_4+r_5}dt \nonumber \\
=&\sum_{\Sigma}{r \choose r_1}{r-r_1 \choose r_2}{r-r_1-r_2 \choose r_3}{r-r_1-r_2-r_3 \choose r_4}{r-r_1-r_2-r_3-r_4 \choose r_5} \nonumber \\
&\frac{2^{r_4+r_5+r_6}b^{r_6}}{N^{2(2-2\alpha-4\xi)r_2+2r_3+(2-2\alpha-4\xi)r_4+r_5+(3-2\alpha-4\xi)r_6}}
3^{2r_1+r_4+r_5}\frac{1}{\delta_0(2r_1+r_4+r_5)}|x|^{2r_1+r_4+r_5} \nonumber\\
=&\sum_{\Sigma}{r \choose r_1}{r-r_1 \choose r_2}{r-r_1-r_2 \choose r_3}{r-r_1-r_2-r_3 \choose r_4}{r-r_1-r_2-r_3-r_4 \choose r_5} \nonumber \\
&\frac{2^{r_4+r_5+r_6}3^{2r_1+r_4+r_5}b^{2r_1+r_4+r_5+r_6}}{\delta_0(2r_1+r_4+r_5)N^{2(2-2\alpha-4\xi)r_2+2r_3+(2-2\alpha-4\xi)r_4+r_5+(3-2\alpha-4\xi)r_6}})
||x||^{2r_1+r_4+r_5} \nonumber
\end{align}
where the second inequality is from Lemma C.5 from \citep{Yin_17}.
Therefore, 
\begin{align}
&\mathbb{E}[||X||^{2r}] \nonumber \\
=&\mathbb{E}[-\sum_{y\neq X}q_{X,y}N(g_r(y)-g_r(X)-\nabla g_r(X)\cdot(y-X))] \nonumber \\
\le&\mathbb{E}[\sum_{\Sigma}{r \choose r_1}{r-r_1 \choose r_2}{r-r_1-r_2 \choose r_3}{r-r_1-r_2-r_3 \choose r_4}{r-r_1-r_2-r_3-r_4 \choose r_5} \nonumber \\
&\frac{2^{r_4+r_5+r_6}3^{2r_1+r_4+r_5}b^{2r_1+r_4+r_5+r_6}}{\delta_0(2r_1+r_4+r_5)N^{2(2-2\alpha-4\xi)r_2+2r_3+(2-2\alpha-4\xi)r_4+r_5+(3-2\alpha-4\xi)r_6-1}}||X||^{2r_1+r_4+r_5}\sum_{y\neq X}q_{X,y}] \nonumber \\
\le&_{(a)}\sum_{\Sigma}{r \choose r_1}{r-r_1 \choose r_2}{r-r_1-r_2 \choose r_3}{r-r_1-r_2-r_3 \choose r_4}{r-r_1-r_2-r_3-r_4 \choose r_5} \nonumber \\
&\frac{2^{r_4+r_5+r_6}3^{2r_1+r_4+r_5}b^{2r_1+r_4+r_5+r_6}}{\delta_0(2r_1+r_4+r_5)N^{2(2-2\alpha-4\xi)r_2+2r_3+(2-2\alpha-4\xi)r_4+r_5+(3-2\alpha-4\xi)r_6-1}}\mathbb{E}[||X||^{2(r_1+\frac{r_4+r_5}{2})}] \nonumber \\
\le&_{(b)}\sum_{\Sigma_1}{r \choose r_1}{r-r_1 \choose r_2}{r-r_1-r_2 \choose r_3}{r-r_1-r_2-r_3 \choose r_4}{r-r_1-r_2-r_3-r_4 \choose r_5} \nonumber \\
&\frac{2^{r_4+r_5+r_6}3^{2r_1+r_4+r_5}b^{2r_1+r_4+r_5+r_6}}{\delta_0(2r_1+r_4+r_5)N^{2(2-2\alpha-4\xi)r_2+2r_3+(2-2\alpha-4\xi)r_4+r_5+(3-2\alpha-4\xi)r_6-1}}\frac{1}{N^{(r_1+\frac{r_4+r_5}{2})(1-4\alpha-7\xi)}} \nonumber\\
=&\sum_{\Sigma_1}{r \choose r_1}{r-r_1 \choose r_2}{r-r_1-r_2 \choose r_3}{r-r_1-r_2-r_3 \choose r_4}{r-r_1-r_2-r_3-r_4 \choose r_5} \nonumber \\
&\frac{2^{r_4+r_5+r_6}3^{2r_1+r_4+r_5}b^{2r_1+r_4+r_5+r_6}}{\delta_0(2r_1+r_4+r_5)N^{(1-4\alpha-7\xi)r_1+2(2-2\alpha-4\xi)r_2+2r_3+(2.5-4\alpha-7.5\xi)r_4+(1.5-2\alpha-3.5\xi)r_5+(3-2\alpha-4\xi)r_6-1}} \nonumber\\
\le&_{(c)}\sum_{\Sigma_1}{r \choose r_1}{r-r_1 \choose r_2}{r-r_1-r_2 \choose r_3}{r-r_1-r_2-r_3 \choose r_4}{r-r_1-r_2-r_3-r_4 \choose r_5} \nonumber \\
&\frac{12\times2^{r_4+r_5+r_6}3^{2r_1+r_4+r_5}b^{2r_1+r_4+r_5+r_6}}{\gamma(2r_1+r_4+r_5)N^{(1-4\alpha-7\xi)r+(3-\xi)r_2+(1+4\alpha+7\xi)r_3+(1.5-0.5\xi)r_4+(0.5+2\alpha+3.5\xi)r_5+(2+2\alpha-11\xi)r_6-1-2\alpha-2\xi}} \nonumber \\
\le &_{(d)}\frac{C\log^{2r-2} N}{N^{(1-4\alpha-7\xi)r+2\alpha+5\xi}}\le\frac{1}{N^{(1-4\alpha-7\xi)r}} 
\label{eq: sigma_1}
\end{align}
where 
\begin{align}
    \Sigma_1=\Sigma\setminus \{r_1=r-1, r_4=1, \text{ and } r_i=0,\quad i=2,3,5,6\}. \nonumber
\end{align}
For the inequality (a), we used the fact that $\sum_{y\neq x}q_{x,y}\le 2$. For the inequality (b), by mathematical induction, we assumed that $$\mathbb{E}[||X||^{2(r-1)}]\le \frac{1}{N^{(1-4\alpha-7\xi)(r-1)}},$$ and by Lyapunov inequality we have $$\mathbb{E}[||X||^{2(r_1+\frac{r_4+r_5}{2})}]\le(\mathbb{E}[||X||^{2(r-1)}])^{\frac{r_1+\frac{r_4+r_5}{2}}{r-1}}.$$ Note that this Lyapunov inequality only holds for the combination set $\Sigma_1$.

For the inequality (c), we applied the result from Lemma C.4 in \citep{Yin_17}. For
the inequality (d), it holds because order-wise the smallest combination of $\{r_i, i=1,\cdots,6\}$ in the set $\Sigma_1$ is when $r_1=r-2,r_5=2$ and $r_i=0$ for $i=2,3,4,6$ and $C>0$ is large enough. And the last inequality holds for sufficiently large $N$.

For $r_1=r-1, r_4=1$ and $r_i=0$ for $i=2,3,5,6$ and this combination has the highest order, then we have the following analysis. By inequality (a), we have
\begin{align}
    E[||X||^{2r}]
    &\le C_1\frac{6\times b^{2r-1}}{\delta_0(2r-1)N^{1-2\alpha-4\xi}}E[||X||^{2r-1}] \nonumber \\
    &\le C_1\frac{6b^{2r-1}\cdot 12N^{2\alpha+3\xi}}{(2r-1)\gamma N^{1-2\alpha-4\xi}}
    E[||X||^{2r-1}] \nonumber \\
    &\le \frac{C_2}{N^{1-4\alpha-8\xi}} E[||X||^{2r-1}],
    \label{eq: E_2r}
\end{align}
where there exists $C_1>0$ and $C_2>0$ for the inequalities to hold.
Again by Lyapunov inequality, we have
\begin{align}
    E[||X||^{2r-1}]\le E[||X||^{2r}]^{\frac{2r-1}{2r}}. 
    \label{eq: Lya}
\end{align}
By combining \eqref{eq: E_2r} and \eqref{eq: Lya}, we have
\begin{align}
     E[||X||^{2r}]\le \frac{C_2}{N^{1-4\alpha-8\xi}} E[||X||^{2r-1}]
     \le \frac{C_2}{N^{1-4\alpha-8\xi}}E[||X||^{2r}]^{\frac{2r-1}{2r}}. \nonumber
\end{align}
As a result, we have
\begin{align}
E[||X||^{2r}]\le\frac{C}{N^{(1-4\alpha-8\xi)\cdot 2r}}\le \frac{1}{N^{(1-4\alpha-7\xi)r}},
\label{eq: another_case}
\end{align}
where the second inequality holds for a sufficiently large $N$. 
When $r_1=r_4=r_5=0$, we have similar analysis based on the fact that $\int_{0}^{\infty}|e(t)|dt\le\frac{1}{N^{2-4\alpha-5\xi}}$.

Therefore, \eqref{eq: hig_mom_bou} holds for all $r\in \mathbb{N}$, by mathematical induction.
\Halmos
\endproof

%%%%%%%%%%%%%%%%%%%%%%%%%%%%%%%%%%%%%%%%%%%%%%%%%%%%%%%%%%%%%%%%%%%%%%%%%%%%
%% Acknowledgments
%%%%%%%%%%%%%%%%%%%%%%%%%%%%%%%%%%%%%%%%%%%%%%%%%%%%%%%%%%%%%%%%%%%%%%%%%%%%

\section*{Acknowledgments}

The authors are very grateful to Nicolas Gast for his valuable comments.
This work was supported in part by NSF ECCS 1739344, CNS 2002608,
and CNS 2001687.

%%%%%%%%%%%%%%%%%%%%%%%%%%%%%%%%%%%%%%%%%%%%%%%%%%%%%%%%%%%%%%%%%%%%%%%%%%%%
%% References
%%%%%%%%%%%%%%%%%%%%%%%%%%%%%%%%%%%%%%%%%%%%%%%%%%%%%%%%%%%%%%%%%%%%%%%%%%%%

\bibliographystyle{plainnat}
\bibliography{inlab-refs}

%%%%%%%%%%%%%%%%%%%%%%%%%%%%%%%%%%%%%%%%%%%%%%%%%%%%%%%%%%%%%%%%%%%%%%%%%%%%

\end{document}